\documentclass[aps,twocolumn,superscriptaddress,nofootinbib,prb]{revtex4-2}

\usepackage{graphicx}% Include figure files
\usepackage{amssymb,amsmath,amsfonts,bm}
\usepackage[normalem]{ulem}
\usepackage{hyperref}
\usepackage[dvipsnames]{xcolor}
\hypersetup{urlcolor=RoyalBlue, colorlinks=true,citecolor=RoyalBlue,linkcolor=RoyalBlue} 
\usepackage{physics}
\usepackage[final]{showkeys}
\usepackage{float}

\begin{document}
\title{Confinement-deconfinement transitions in the dimer-loop models on two-dimensional non-bipartite lattices}

\author{Jay Pandey}
\affiliation{Department of Theoretical Physics, Tata Institute of Fundamental Research, 1 Homi Bhabha Road, Mumbai 400005}
\author{Souvik Kundu}%
\affiliation{International Centre for Theoretical Sciences, Tata Institute of Fundamental Research, Bengaluru 560089, India}
\author{Kedar Damle}
\affiliation{Department of Theoretical Physics, Tata Institute of Fundamental Research, 1 Homi Bhabha Road, Mumbai 400005}

\date{\today}

\begin{abstract}
Recent work~\cite{Kundu_Damle_PRX_2025} has demonstrated that a class of $S=1$ kagome lattice antiferromagnets with competing anisotropies can exhibit unusual low-temperature behavior that is best described in terms of the statistical mechanics of an interesting interpolation between the fully-packed dimer model and the fully-packed loop model on the bipartite honeycomb lattice. On this and other bipartite lattices in two and three dimension, such fully-packed dimer-loop models generically have two distinct power-law correlated Coulomb phases, both of which admit a low-energy description in terms of a fluctuating divergence-free polarization vector field~\cite{Kundu_Damle_PRX_2025,Kundu_Damle3Dpreprint2025}. The transition between these phases corresponds to a confinement transition for the $Z_2$ part of the corresponding $U(1)$ flux~\cite{Kundu_Damle_PRX_2025,Kundu_Damle3Dpreprint2025}. Here we demonstrate that a variant of such a topological confinement transition  can also exist in dimer-loop models on non-bipartite lattices such as the kagome and triangular lattices although the low-energy description does not feature any conserved $U(1)$ flux. In these non-bipartite dimer-loop models, both phases are short-range correlated liquids, and the topological distinction between them is in terms of a $Z_2$-valued winding number instead of the flux of a divergence-free polarization field. In spite of this key difference, this transition appears to be in the same Ising universality class as the analogous transition of the bipartite dimer-loop models. These kagome and triangular lattice dimer-loop models correspond respectively to the low-energy limit of a class of $S=1$ antiferromagnets with competing anisotropies on the ruby lattice comprising a corner-sharing arrangement of fully-frustrated squares and a lattice of corner-sharing fully-frustrated hexagons.

\end{abstract}

%Use showkeys class option if keyword
%display desired
\maketitle

\section{Introduction}
When antiferromagnetic interactions between magnetic ions compete due to lattice geometry, the magnetic moments tend to evade ordering down to very low temperature. At a classical level, this is because there are a macroscopically large number of symmetry-inequivalent configurations that minimize the interaction energy. This entropy-dominated physics of geometric frustration opens up a large low-temperature window in which physical properties are dominated by low-energy fluctuations within the minimally-frustrated configuration space. Perhaps the best known examples of these are kagome antiferromagnets in two dimensions and pyrochlore antiferromagnets in three dimensions~\cite{Mila_Mendels_Lacroix_2011book}.

Recent work~\cite{Kundu_Damle_PRX_2025} has demonstrated that a class of $S=1$ kagome lattice antiferromagnets with competing anisotropies can exhibit unusual low-temperature behavior that is best described in terms of the statistical mechanics of an interesting interpolation between the fully-packed dimer model and the fully-packed loop model on the bipartite honeycomb lattice. 
On this and other bipartite lattices in two and three dimension~\cite{Kundu_Damle_PRX_2025,Kundu_Damle3Dpreprint2025}, such fully-packed dimer-loop models were found to generically have two distinct power-law correlated Coulomb phases~\cite{Henley_Coulombphasesreview2010}, both of which admit a low-energy description in terms of a fluctuating divergence-free polarization vector field. The transition between these phases corresponds to a confinement transition for the $Z_2$ part of the corresponding $U(1)$ flux. In related work, similar flux confinement transitions have been argued to exist in pyrochlore antiferromagnets with such competing anisotropies~\cite{Pandey_Damle_Spin1preprint2025,Pandey_Kundu_Damle_Spin3by2preprint2026}. 

The statistical mechanics of the fully-packed dimer model and the full-packed loop models on bipartite lattices has been the subject of sustained theoretical development over many 
decades\cite{Baxter_1989,Youngblood_Axe_1981,Blote_Nienhuis_1994_PRL,Batchelor_etal_1996,
Fendley_etal_2002_PRB,Verrsen_Vishwanath_2022,Misguich_etal_2002_PRL,Wilkins_Powell_PRE_2023,
Henley_Coulombphasesreview2010,KASTELEYN_1961_1209,Fisher_1961,Fisher_Stephenson_1963}.
Both models feature dipolar power-law correlations at long distances, described by an effective field theory for a fluctuating divergence-free polarization field~\cite{Henley_Coulombphasesreview2010}. The bipartite dimer-loop models studied recently~\cite{Kundu_Damle_PRX_2025,Kundu_Damle3Dpreprint2025} interpolate in a natural way between these two well-understood limiting cases. Like the dimer and the loop models, they also admit a description in terms of the same effective field theory for a fluctuating divergence-free polarization field~\cite{Kundu_Damle_PRX_2025,Kundu_Damle3Dpreprint2025}. On this basis, one might have imagined that properties of the dimer-loop model interpolate continuously between the two limiting cases as one tunes the relative fugacity $w$ of dimers. However, in actual fact, these dimer-loop models have a phase transition at which a topological property displays sharp threshold behavior~\cite{Kundu_Damle_PRX_2025,Kundu_Damle3Dpreprint2025}. Thus, such dimer-loop models are interesting from a purely theoretical point of view as well, quite independent of the possibility that they describe interesting low-temperature phenomena in some frustrated magnets.

Here we demonstrate that a variant of such a topological confinement transition  can  also exist in dimer-loop models on non-bipartite lattices such as the kagome and triangular lattices although there is no notion of a divergence-free $U(1)$ flux. These models also provide the low-energy description of a class of $S=1$ antiferromagnets with competing anisotropies on the ruby lattice comprising a corner-sharing arrangement of fully-frustrated squares and a lattice of corner-sharing fully-frustrated hexagons. In these nonbipartite dimer-loop models, both phases are short-range correlated liquids, and the topological distinction between them is in terms of a $Z_2$-valued winding number instead of the flux of a divergence-free polarization field. In spite of this key difference, this transition appears to be in the same Ising universality class as the analogous transition of the bipartite dimer-loop models~\cite{Kundu_Damle_PRX_2025,Kundu_Damle3Dpreprint2025}. 

This resolves one of the two key questions raised by the work of Ref.~\cite{Kundu_Damle_PRX_2025}, namely, the extent to which the bipartiteness of the lattice is crucial for the physics of this topological transition. The other intriguing question raised by Ref.~\cite{Kundu_Damle_PRX_2025} has to do with the fact that some aspects of the transition have Ising character, and concern the explicit identification of a local Ising field whose long-wavelength properties near the transition match those of the Ising model in its critical regime. This is addressed in parallel work~\cite{Pandey_Damle_DimerLoopIsingorderarXiv2026}, and we will have occasion to use one of the insights from that work in the analysis of our results on test-defect correlators at the nonbipartite transition studied here.

The rest of this paper is organized as follows: In Sec.~\ref{sec:Model} we define the models we study on the kagome and triangular lattices, provide some additional background about previous results in the bipartite case and spell out the connection to spin $S=1$ antiferromagnets on the ruby lattice comprising fully-frustrated corner-sharing squares and a lattice of fully-frustrated corner-sharing hexagons. We also define the $Z_2$-valued winding number invariants that help us characterise the unusual transition we identify. In Sec.~\ref{sec:Methods}, we describe our numerical methods and the observables we focus on. Sec.~\ref{sec:Results} is devoted to our results, and we discuss some outstanding questions and directions for future work in Sec.~\ref{sec:Discussion}

\begin{figure}[htbp]
\centering
\begin{tabular}{cc}
\includegraphics[width=0.49\linewidth]{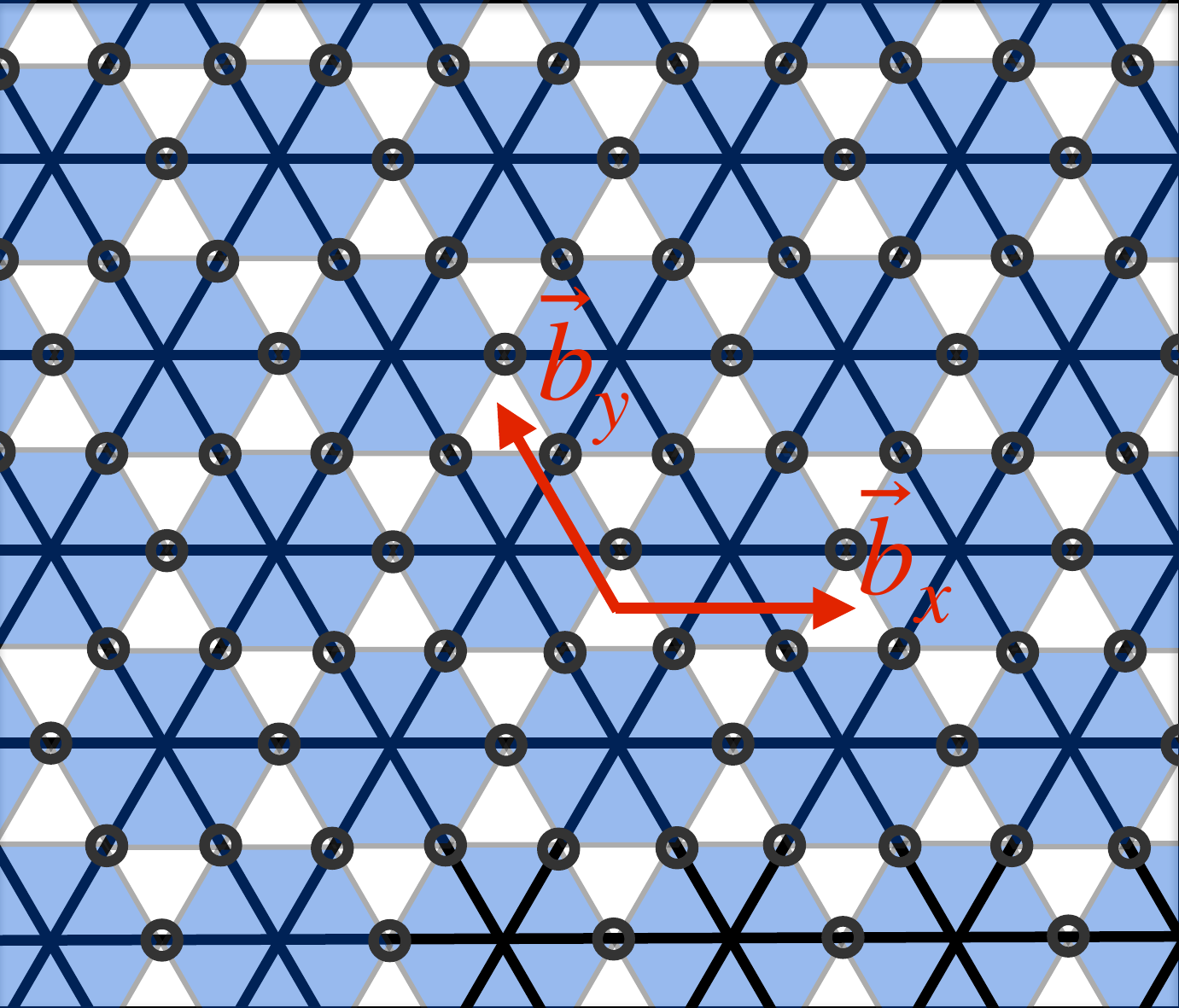} &
\includegraphics[width=0.49\linewidth]{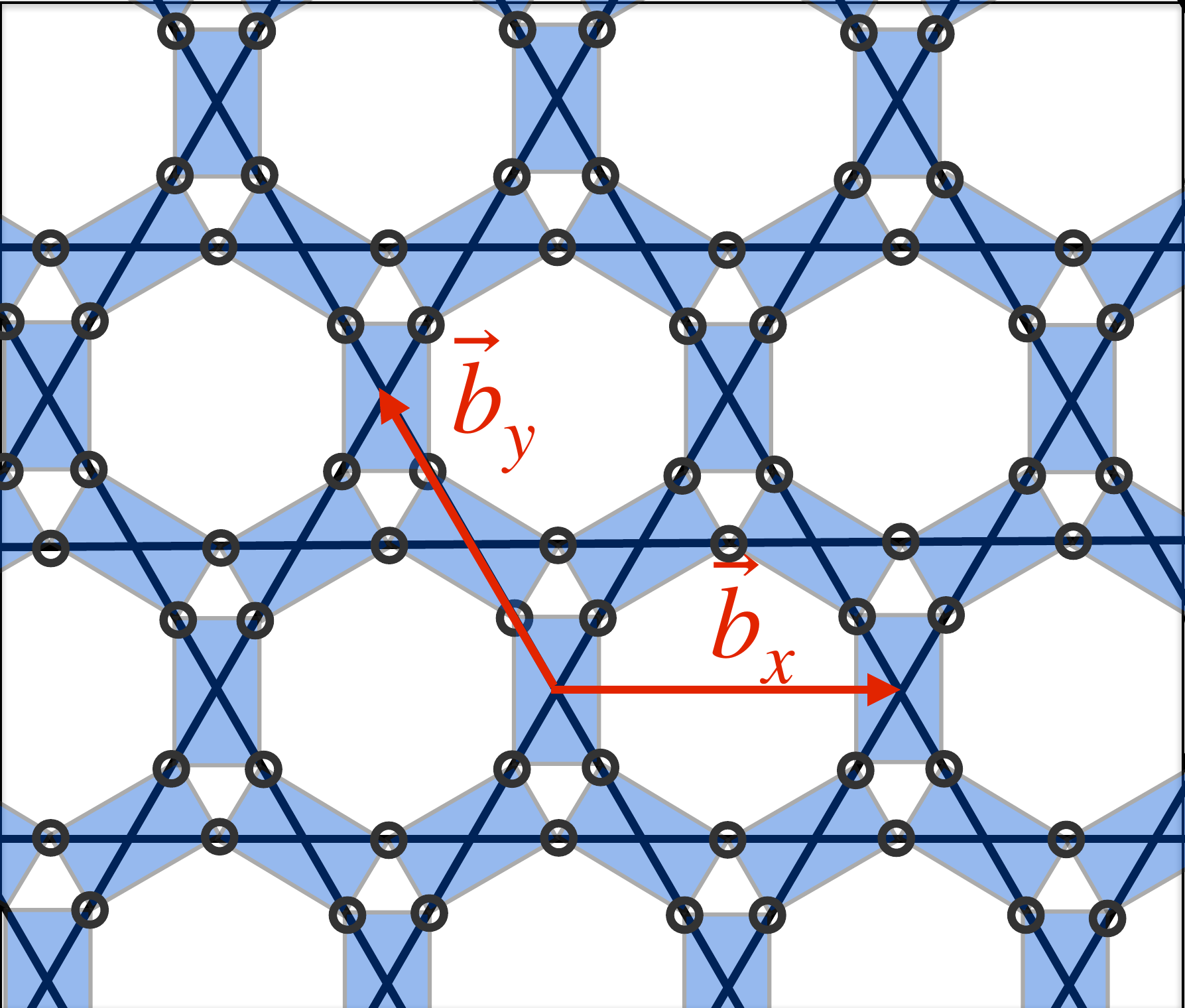} \\
(a) & (b)
\end{tabular}

\caption{The dimer-loop model Eq.~\ref{eq:partition_function} studied here is defined on the triangular and kagome lattices. (a) The links (black lines) of the triangular lattice host the spins of a $S=1$ antiferromagnet defined on a lattice of fully-frustrated corner-sharing hexagons (see Eq.~\ref{eq:frustratedmagnet}).  
(b) The links (black lines) of the kagome lattice host the spins of a $S=1$ antiferromagnet defined on a lattice of fully-frustrated corner-sharing squares that form the ruby lattice (see Eq.~\ref{eq:frustratedmagnet}).}
\label{fig:side_by_side_lattices}
\end{figure}
\section{Motivation and model}
\label{sec:Model}

\subsection{Definition}
\label{subsec:Model}
\begin{figure}
\includegraphics[width=\linewidth]{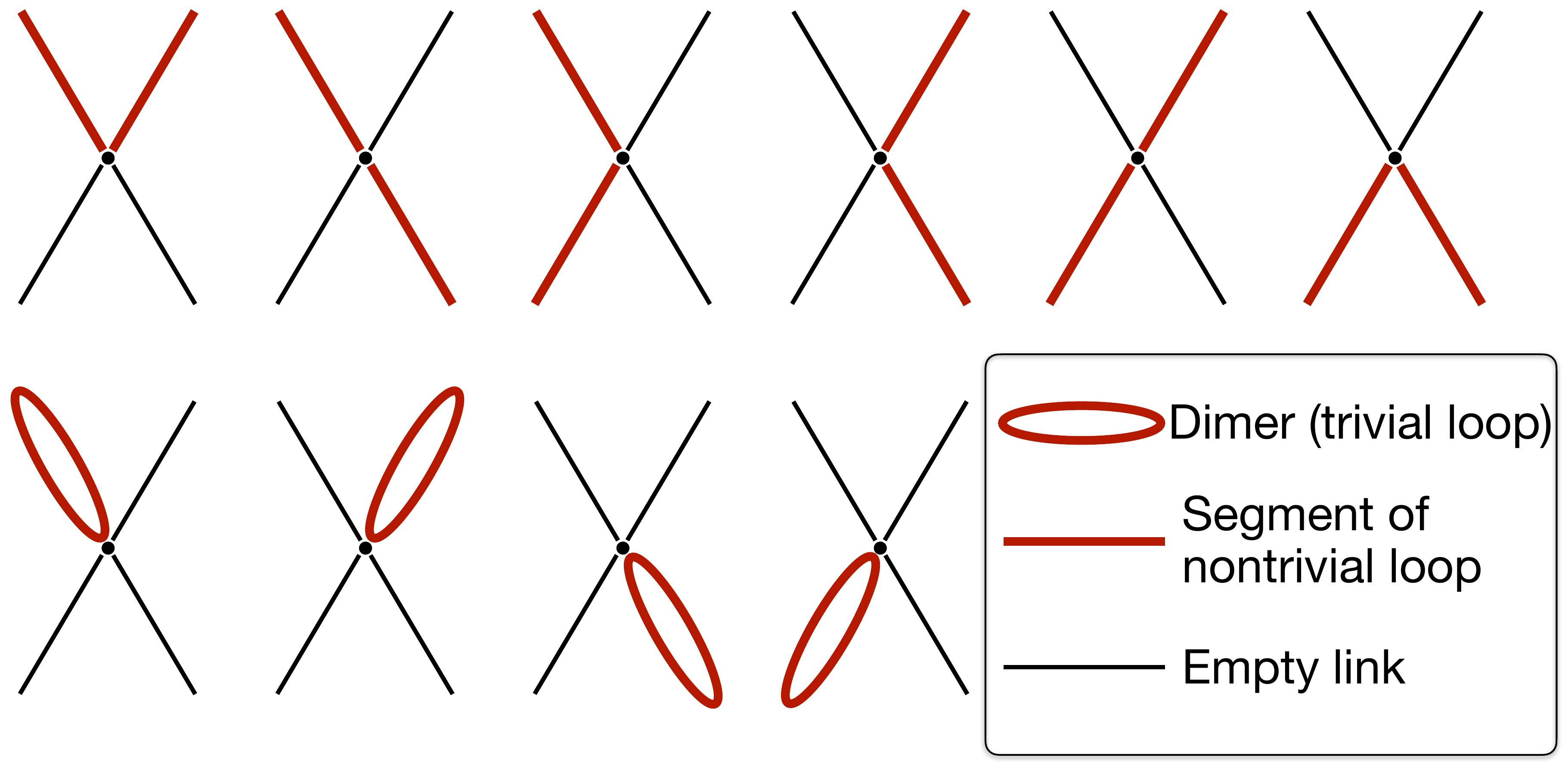}
\caption{A schematic of the various ways in which a vertex of the kagome lattice can be touched by either a dimer (trivial loop, length $s =2$) or a non-trivial loop (of length $s\geq 2$) in a valid fully packed configuration that contributes to the partition function $Z(w)$. The set of possibilities at a vertex of the triangular lattice can be enumerated in a similar way.} \label{fig:KagomeVertexConfigs}
\end{figure}

We focus on the properties of the fully-packed dimer-loop model on two prototypical planar non-bipartite lattices, the triangular lattice and the kagome lattice (Fig.~\ref{fig:side_by_side_lattices}). Here, dimers are hard rods that live on a link of the lattice and touch the two vertices connected by that link.  By loops, we mean non-intersecting non-backtracking simple loops of length greater than $2$, that live on links of the lattice. The dimers and loops jointly obey a hard-core constraint: If a vertex is touched by a dimer, it cannot have a loop passing through it, and a vertex not touched by a dimer can have at most one loop passing through it. Our focus here is on the fully-packed case, in which configurations contributing to the partition function obey an additional full-packing constraint: Every vertex must either be touched by a single dimer or have a single loop segment passing through it. By way of a concrete example, the different ways in which the full-packing constraint can be obeyed at a given vertex  of the kagome lattice are shown in Fig.~(\ref{fig:KagomeVertexConfigs}). The corresponding set of possibilities for the triangular lattice can be enumerated in the same way.

\begin{figure}
\includegraphics[width=\linewidth]{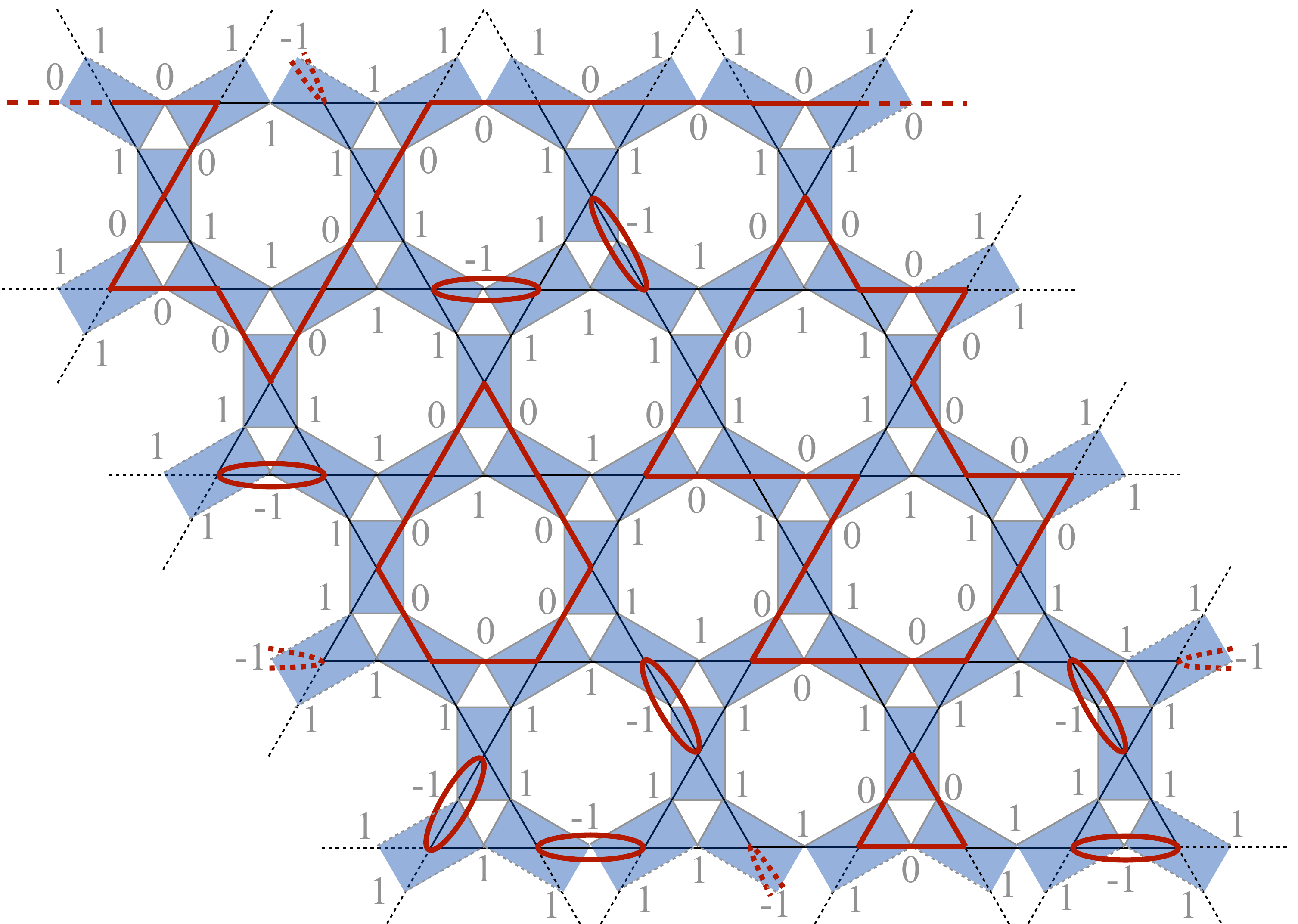}
\caption{Example of an allowed configuration of the fully packed dimer-loop model on the kagome lattice with periodic boundary conditions (periodic boundary conditions indicated by dotted lines at the boundaries). The dimers have been drawn as ``trivial loops'' connecting two adjacent vertices, traversing the link between them in both directions. Also shown is the corresponding configuration of spins $S^z_i$ on sites $i$ of the ruby lattice made up of corner-sharing fully-frustrated squares (see Eq.~\eqref{eq:frustratedmagnet}).}
\label{fig:ExampleConfigKagome}
\end{figure}

We consider the statistical mechanics of such systems, with the statistical weight $\mathcal{W}$ of any fully-packed configuration $C$ of dimers and loops defined to be $\mathcal{W}(C) = w^{N_d(C)}$, where $N_d(C)$ is the number of dimers in the configuration $C$. An example of a legal configuration on the kagome lattice is shown in Fig.~(\ref{fig:ExampleConfigKagome}). Thus, the dimers have a fugacity $w$, while loops have unit fugacity. The partition function $Z(w)$ is given by:
\begin{equation}
Z(w) = \sum_C w^{N_d(C)} \; .
\label{eq:partition_function}
\end{equation}  
$Z(w)$ interpolates continuously between the partition function of the fully-packed loop model ($w=0$) and the partition function of the fully-packed dimer model $(w=\infty)$.

We define the length $s$ of a loop  to be the number of vertices it passes through. 
Following Ref.~\cite{Kundu_Damle_PRX_2025}, we find it useful for some purposes to view a dimer as a trivial or degenerate loop that goes from a vertex to a neighbor along a link and returns to the starting vertex along the same link.  Dimers are therefore assigned a length of $s=2$. When we wish to emphasize this point of view, we refer to loops of length $s>2$ as ``nontrivial" loops, and dimers as ``trivial'' loops. In this language, length $s=2$ trivial loops have fugacity $w$ and all other loops of length $s >2$ have unit fugacity.  Note that this assignment leads to the sum rule:
\begin{equation}
\sum_{j}s_j = N_{\rm sites}
\end{equation} 
where the sum over $j$ is over {\em all loops} (both trivial loops and nontrivial loops) in a fully-packed configuration and $N_{\rm sites}$ is the number of sites of the underlying lattice.

This assignment of length $s=2$ to dimers emphasizes the distinction between dimers and an open string of length $s=1$. In the honeycomb lattice case studied earlier~\cite{Kundu_Damle_PRX_2025},  certain temperature and field regimes of the corresponding kagome lattice magnet are expected to be modeled by a generalization of Eq.~\ref{eq:partition_function} that allows additional configurations that contain some open strings of various lengths (including length $s=1$ strings),  and thereby violate the full-packing constraint at some vertices~\cite{Kundu_Damle3Dpreprint2025}. In the present work, we focus on the non-bipartite fully-packed case, leaving such generalizations aside for future work.

\subsection{Connection to frustrated $S=1$ antiferromagnets with competing anisotropies}
As mentioned earlier, the dimer-loop models we study also provide a low-energy description of a class of $S=1$ antiferromagnets with competing anisotropies on the ruby lattice comprising fully-frustrated corner-sharing squares and a lattice of fully-frustrated corner-sharing hexagons; see Fig.~\ref{fig:side_by_side_lattices} for a pictorial depiction of the connection. Here, we provide a brief summary of this connection, although our focus is the statistical mechanics of the dimer-loop model itself. Consider the spin $S=1$ Hamiltonian
\begin{eqnarray}
H &=& \large\sum_{t} \large \sum_{\langle ij \rangle \in t} \left(J_zS^z_iS^z_j+J_{\perp}(S^x_iS^x_j+S^y_iS^y_j)\right)_t \nonumber \\
&& + \sum_i\Delta (S^z_i)^2-B\sum_i S^z_i \; ,
\label{eq:frustratedmagnet}
\end{eqnarray}
where $\langle i j \rangle \in t$ denotes all pairs of sites belonging to the same square (hexagon) $t$ of the ruby lattice of corner-sharing fully-frustrated squares (corner-sharing fully-frustrated hexagons) shown in Fig.~(\ref{fig:side_by_side_lattices}). We focus on the regime in which the quantum fluctuations induced by $J_{\perp}$ are negligible compared to the classical energetics controlled by $J_z$, $B$ and $\Delta$, which are all comparable to each other and represent the largest energy scales in the problem:
\begin{eqnarray}
&&J_z = J \; \; > \; \; 0 \; ,\nonumber \\
&&\Delta = J +\mu  \; ,  \nonumber \\
&&B  = 2 \alpha J + \delta B\; ,\nonumber \\
&&\mu , \delta B  \ll  J \; ,
\end{eqnarray}
where $\alpha = 2$ for fully-frustrated corner-sharing squares and $\alpha = 4$ for fully-frustrated corner-sharing hexagons. For magnetic fields in this range, the magnet on the ruby lattice of corner-sharing fully-frustrated squares (lattice of corner-sharing fully-frustrated hexagons) is expected at low temperature to develop a $1/2$-magnetization ($2/3$-magnetization) plateau. Further, due to the separation of energy scales, there is a sizeable low-temperature window in which the behavior on this plateau is controlled by the effect of thermal fluctuations acting within the low-energy subset of configurations that correspond to the plateau. 

These low-energy configurations have the following simple pictorial description in terms of the dimer-loop model on the corresponding kagome or triangular lattice: Represent each $S^z = -1$ by a dimer on the corresponding link of the  kagome or triangular lattice, each $S^z = 0$ by a segment of a nontrivial loop on the corresponding link, and each $S^z=+1$ by an empty or unoccupied link. In this representation (see Fig.~\ref{fig:ExampleConfigKagome}), thermal fluctuations of the spins at low temperature correspond exactly to statistical mechanics of the fully-packed dimer-loop model with fugacity
$w = \exp(-2\mu/T)$ for each dimer.

\subsection{$Z_2$ valued winding number invariant}
\label{subsec:unsigned_invariant}

\begin{figure}[htbp]
\centering
\begin{tabular}{c}
\includegraphics[width=\linewidth]{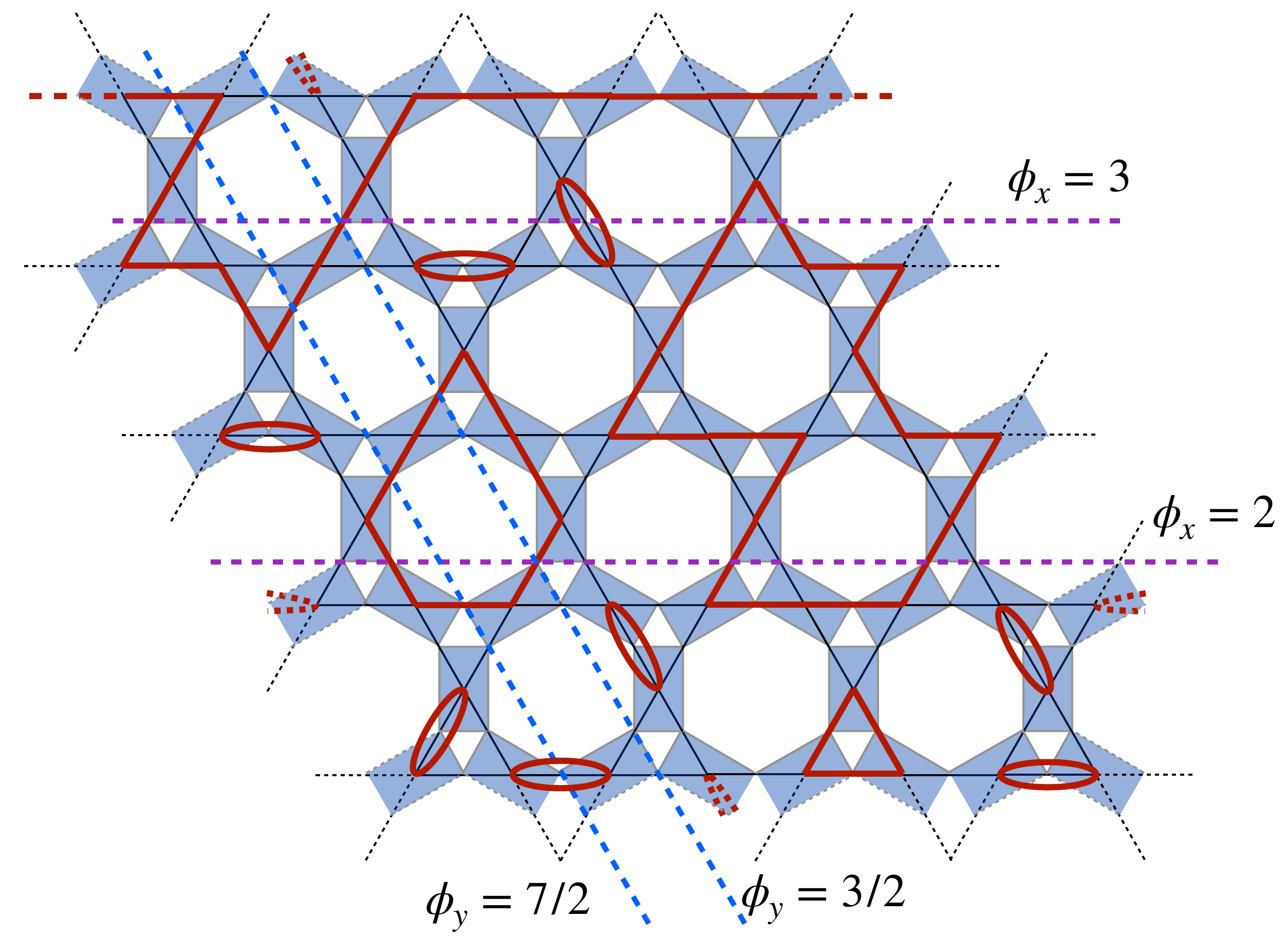} \\
(a) \\[0.5cm]
\includegraphics[width=\linewidth]{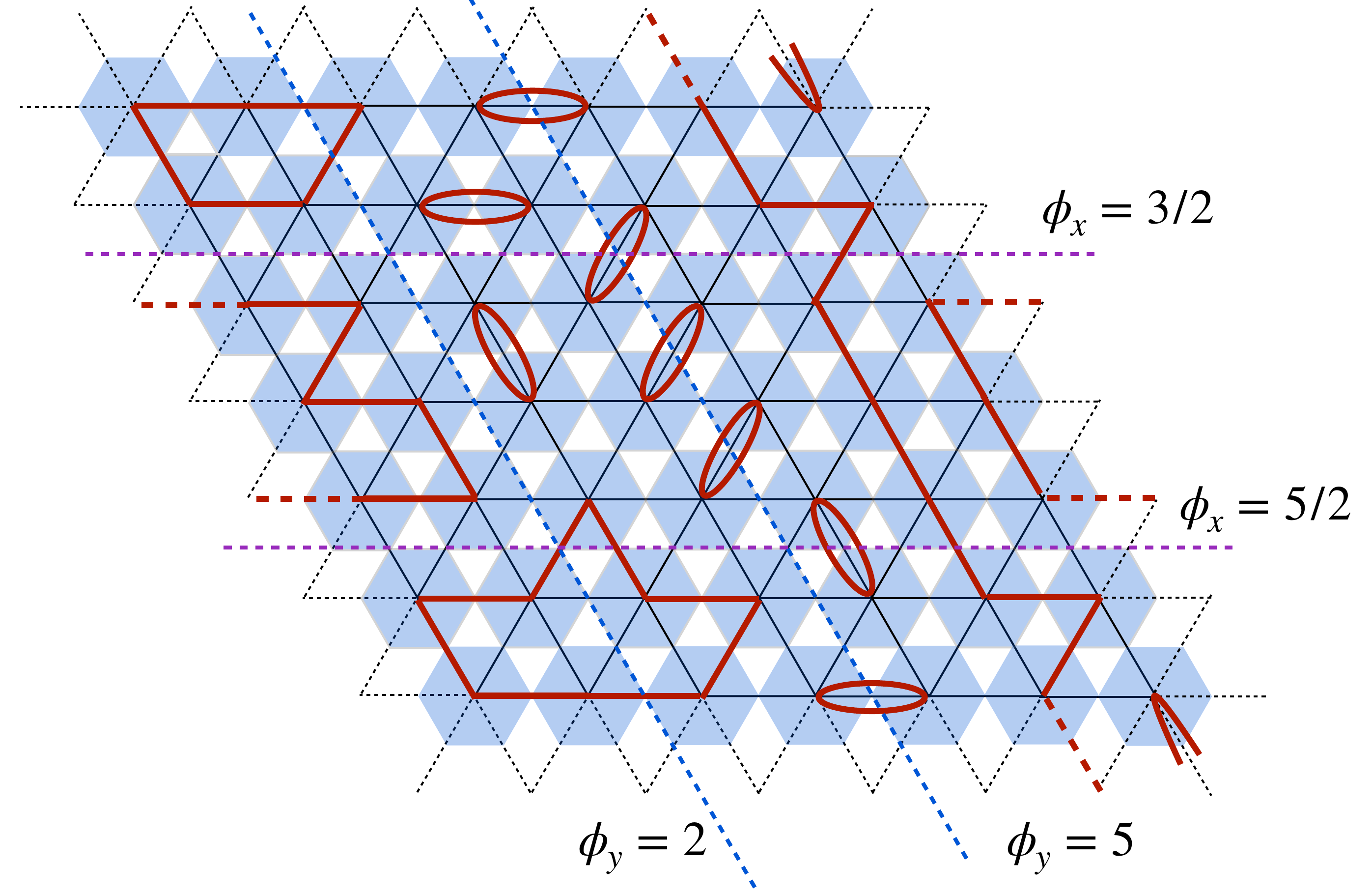} \\
(b)
\end{tabular}

\caption{(a) and (b) Examples of computations of $(\phi_x(c),\phi_y(c))$, defined in Eq.~\eqref{eq:WindingNumberDef}, for a pair of parallel cuts $c$ (denoted by dashed blue (purple) lines for vertical (horizontal) cuts) of a given dimer-loop configuration on the kagome and the triangular lattices, respectively. These examples illustrate two things: First, they show that these $\phi_x$ and $\phi_y$ variables have an integer ambiguity. Second, they illustrate the fact that the $Z_2$-valued winding numbers $W_x$ and $W_y$, which are only sensitive to the presence of a fractional part of $1/2$ in the computed values, are well-defined properties of a given configuration.}
\label{fig:UnsignedInvariant}
\end{figure}

On planar bipartite lattices such as the square and the honeycomb lattice, configurations of the fully-packed dimer-loop model map to configurations of a divergence-free polarization field on links of the lattice~\cite{Kundu_Damle_PRX_2025,Kundu_Damle3Dpreprint2025,Henley_Coulombphasesreview2010}. Violations of the full-packing constraint correspond to the creation of charges that serve as sources of this polarization field. In the fully-packed case, one can solve the zero-divergence constraint in terms of a height field defined on vertices of the dual lattice in two dimensions (or a vector potential defined on links of the dual lattice in three dimensions).
This is not possible for a non-bipartite dimer-loop model. Indeed, on non-bipartite lattices, there is no consistent global convention that can convert the scalar occupation number variables (indicating presence of a dimer or loop segment) on links of the lattice to vector variables that can play the role of a polarization field. As a result, there is no consistent notion of a flux or flux sectors in such systems. 

The topological distinction between the two Coulomb phases of dimer-loop models on bipartite lattices was originally characterized in terms of flux sectors that survive the thermodynamic limit with periodic boundary conditions and test-charge correlators that are independent of boundary conditions~\cite{Kundu_Damle_PRX_2025,Kundu_Damle3Dpreprint2025}. One might therefore wonder if two topologically-distinct phases are at all possible on non-bipartite lattices.
In this context, it is useful to recall that fully-packed configurations of dimers on the triangular lattice with periodic boundary conditions can be classified according to a different topological invariant~\cite{Moessner_QuantumDimerModelsReview2008}. This consists of a $Z_2$-valued winding number for each periodic direction. 

Here, we generalize this definition to include loops and show that our generalization yields a consistent and well-defined invariant so long as the configurations are fully packed. The same generalization also works on the kagome lattice: For any fully packed configuration of dimers and loops on the kagome or triangular lattice with periodic boundary conditions, we define an occupation number variable $n_{\langle r r'\rangle}$  that takes on the following values on links $\langle r r' \rangle$ of the kagome or triangular lattice: 
\begin{eqnarray}
n_{\langle r r'\rangle} &=& 0 \; {\rm if} \; {\rm link} \, \langle r r'\rangle \; {\rm empty} \nonumber \\
n_{\langle r r'\rangle} &=& \frac{1}{2} \; {\rm if} \; \langle r r'\rangle \; {\rm hosts} \; {\rm loop} \; {\rm  segment} \nonumber \\
n_{\langle r r'\rangle} &=& 1\; {\rm if} \; \langle r r'\rangle \; {\rm hosts} \; {\rm dimer}
\end{eqnarray}
We now take a cut along one of the two principal axes. For a given cut $c$ along the principal axis $i$, $\phi_i(c)$ is the sum of these occupation number variables corresponding to the links crossed by this cut. In this sum, dimers (loop segments) have weight $1$ $(1/2)$, so we have
\begin{equation}
\phi_i(c) = n_{\rm dimer}(c) + \frac{1}{2}n_{\rm loop} (c), \label{eq:WindingNumberDef}
\end{equation}
where $c$ is the cut in question, $i=x,y$,  and $n_{\rm dimer}(c)$ ($n_{\rm loop}(c)$) are the number of links crossed by $c$ which host dimers (non-trivial loops). Consequently, $\phi_i(c)$ takes either integer values or has fractional part of $1/2$. This definition is illustrated in Fig.~(\ref{fig:UnsignedInvariant}). Although the actual half-integer values of $\phi_x$ and $\phi_y$ can be different when evaluated along different parallel cuts of a configuration, all these evaluations are guaranteed to agree on whether the quantity is an integer or has a fractional part of $1/2$. We denote this invariant information content to be the $Z_2$ winding number $W_i$ along the principal axis $i$, with the convention that $W_i = 0$ indicates that $\phi_i(c)$ is an integer for all parallel cuts $c$ along axis $i$ while $W_i=1$ indicates the existence of a fractional part of $1/2$ for all such $\phi_i(c)$.

The proof of this is straightforward: Dimers contribute integers to any $\phi_i(c)$ and can therefore lead to at most an integer ambiguity in these variables. As for loops, consider a principal axis $i$ and a cut parallel to and just below it. Let $l$ be a loop segment that crosses this cut. If $l$ is part of a contractible loop, it will necessarily cross the cut an even number of times, and hence lead to at most an integer ambiguity relative to the value obtained at any other cut parallel to $c$. If it is a non-contractible loop winding around the torus in a direction linearly independent of the orientation of the cut, then it will necessarily contribute a fractional part of $1/2$ to all cuts parallel to $c$. This shows that the variables $W_i$ are well-defined. 

We emphasize that this definition of $W_i$, and this argument that shows they are well-defined, both differ from the analogous ideas for the triangular lattice dimer model~\cite{Moessner_QuantumDimerModelsReview2008} in an important way: In that case, the invariant distinction is in terms of whether the winding number is even or odd, and requires that the linear dimensions of the sample be even. Since we too work with $L_x$ and $L_y$ even, one might be tempted to therefore think that our generalization yields a $Z_4$-valued invariant, but this is false. The only well-defined distinction for fully-packed dimer-loop configurations has to do with whether the winding number has a fractional part of $1/2$ or not.  This is illustrated for example by the values of $\phi_x(c)$ evaluated along two parallel cuts for the fully-packed kagome lattice configuration shown in Fig.~(\ref{fig:UnsignedInvariant}) (a) and $\phi_y(c)$ evaluated for two parallel cuts for the fully-packed triangular lattice configuration shown in  Fig.~(\ref{fig:UnsignedInvariant}) (b).

\section{Monte Carlo algorithm and measurements}
\label{sec:Methods}
We performed Monte Carlo simulations of the dimer-loop model on the triangular and kagome lattices with periodic boundary conditions and linear system sizes $L$ chosen to be multiples of six (as measured by the number of unit cells traversed). Specifically, for the kagome lattice, we use $L=144$, $L=192$, $L=384$, and $L=768$, while for the triangular lattice we use $L=144$, $L=288$, $L=384$, $L=576$, $L=720$, and $L=1440$. On the triangular lattice, each unit cell of course contains only a single site, whereas the kagome lattice unit cell is made up of three sites forming an up-pointing triangle.

Here, we focus on reviewing the algorithm of Ref.~\cite{Kundu_Damle_PRX_2025}, which uses the worm algorithm framework~\cite{Sandvik_Moessner_2006,Alet_etal_PRE_2006} to develop a procedure that efficiently updates bipartite dimer-loop configurations. With this in hand, we describe the key modification needed in our work to account for the non-bipartite nature of the underlying lattices considered here. Although there is no notion of ``vortices'' in the non-bipartite cases considered here, we retain the terminology used earlier~\cite{Kundu_Damle_PRX_2025,Kundu_Damle3Dpreprint2025} to facilitate easy comparison with the corresponding update schemes described in Ref.~\cite{Kundu_Damle_PRX_2025} and Ref.~\cite{Kundu_Damle3Dpreprint2025}.

The half-vortex worm update preserves detailed balance in an extended configuration space in which the worm tail is fixed at a kagome or triangular site, such that the local magnetization constraint on the corresponding fully frustrated simplex of the spin model (a fully-frustrated square or hexagon) is violated in the minimal possible way, i.e., where the total spin satisfies $\sum_{i \in \text{simplex}} S_i^z = \alpha + 1$ ($\alpha = 2$ for the kagome lattice and $\alpha = 4$ for the triangular lattice) giving rise to a static $\delta S^z_{\rm tot} = +1$ defect, while the worm head---carrying an equal and opposite violation $\delta S^z_{\rm tot} = -1$ of the total spin on the corresponding simplex---propagates through the lattice with detailed balance. In the equivalent dimer–loop representation, the worm tail corresponds to the site at which an open string terminates, while the worm head denotes a site from which three loop segments emanate or a loop segment touches a dimer. 

Similarly, the unit-vortex update involves a fixed worm tail at a site where the constraint is violated by two units ({\em i.e.} $\sum_{i \in \text{simplex}} S_i^z = \alpha + 2$ at the corresponding simplex of the frustrated magnet), with the worm head marking the location of the opposite violation with $\delta S^z_{\rm tot} = -2$. By construction, the unit-vortex worm head is not allowed to split into two halves. This is an algorithmic constraint rather than a physical one. In the dimer-loop representation, the worm tail corresponds to a fixed site that is not touched by any loop segment or dimer, whereas the worm head denotes a site touched by two dimers.

This is essentially identical to the algorithm published in Ref.~\cite{Kundu_Damle_PRX_2025}, and the proof of detailed balance given in Ref.~\cite{Kundu_Damle3Dpreprint2025} for this algorithm (and for its generalization to include open strings in the dimer-loop picture) also holds for our case in spite of the fact that there is one new feature introduced by the non-bipartite nature of the lattice. To understand this new feature, and how one can deal with it without violating detailed balance, it is necessary to first review the algorithm of Ref.~\cite{Kundu_Damle_PRX_2025} in more detail.
\subsection{Half-vortex update}

Each update begins with a fully packed configuration of dimers and loops. A site is picked randomly and designated as the first entry point $e_0$. There are two possibilities now. First possibility: if $e_0$ is touched by a dimer, the worm construction is either terminated with probability $1/2$  (without making any change to the configuration), or, a worm update is attempted with probability $1/2$. In the latter case, the site at the opposite end of this dimer is designated as the first pivot site $\pi_0$. Second possibility: if, instead, $e_0$ is part of a nontrivial loop segment, one of the two neighboring sites that lie on this loop is chosen randomly (with probability $1/2$ each) and designated the first pivot site $\pi_0$. After reaching a pivot, the worm construction proceeds through a sequence of steps that follow the procedure that we now describe. 

At the $n$th step, the worm-head arrives at a pivot site $\pi_n$ from an entry site $e_n$.  The local environment of $\pi_n$ is now altered in a way that locally respects detailed balance. Specifically,  with probabilities chosen to satisfy detailed balance, one of the neighbors of $\pi_n$ is picked as the exit site $e_n'$. 

The operational meaning of this is as follows: If $e_n'$ is different from $e_n$, then the link occupation number $n_{\langle \pi_n e_n\rangle}$ is decreased by $1/2$ and the link occupation number $n_{\langle \pi_n e_n'\rangle}$ is increased by $1/2$. Thus, if  link $\langle \pi_n e_n\rangle$ initially hosted a loop segment, this segment is pivoted away from $\langle \pi_n e_n\rangle$ and instead placed on link $\langle \pi_n e_n'\rangle$. If link $\langle \pi_n e_n\rangle$ was initially occupied by a dimer, that dimer is converted into a loop segment and another loop segment is also placed on link $\langle \pi_n e_n'\rangle$.  

In this way, the worm head moves to the exit site $e_n'$. Note that this means that the $\delta S^z_{\rm tot} = -1$ defect is now located at $e_n'$, which is then designated as the entry site $e_{n+1}$ for the next step of the worm motion. If $e_{n'}$ coincides with $e_{n}$, then there is no change to the local environment of $\pi_n$ and $e_{n}$ itself is designated as the entry site $e_{n+1}$ for the next step.

Once $e_{n+1}$ has been thus designated, it remains to specify the protocol to be followed for picking the next pivot site $\pi_{n+1}$. This proceeds as follows: If the process of exiting the pivot $\pi_n$ via exit $e_n'$ creates a loop segment on $\langle \pi_n e_n'\rangle$ and site $e_n'$ was already touched by two other loop segments that were part of a single loop, then we choose one of these two loop segments with probability $1/2$ and designate the site at its other end to be the next pivot $\pi_{n+1}$. If the process of exiting the pivot $\pi_n$ via exit $e_n'$ creates a loop segment on $\langle \pi_n e_n'\rangle$ and site $e_n'$ was already touched by a dimer, then the site at the other end of this dimer is designated to be the next pivot $\pi_{n+1}$. If the process of exiting the pivot $\pi_n$ via exit $e_n'$ converts a preexisting loop segment on $\langle \pi_n e_n'\rangle$ into a dimer on $\langle \pi_n e_n'\rangle$ , then $e_n'$ must necessarily be touched by a single other loop segment that originally belonged to the same loop as this preexisting loop segment. In this case, the site at the other end of this other loop segment is designated as the next pivot $\pi_{n+1}$. In all these cases, we view this next pivot as being reached from entry site $e_{n+1} \equiv e_n'$. 

Finally, if $e_n'$ coincides with $e_0$, then there is no next pivot and the defect at the worm head annihilates the defect at the worm tail to yield a new fully-packed dimer configuration once we increase the link occupation number $n_{\langle \pi_n e_n' \rangle}$ by $1/2$ (which either places a loop segment on the previously empty link $\langle \pi_n e_n' \rangle$ or converts the pre-existing loop segment on $\langle \pi_n e_n' \rangle$ to a dimer on $\langle \pi_n e_n' \rangle$. This new fully-packed dimer-loop configuration is accepted with probability $1$.

Having reviewed the algorithm of  Ref.~\cite{Kundu_Damle_PRX_2025}, we come to the key difference introduced by the non-bipartite nature of the triangular or kagome lattice: Since non-bipartite lattices have odd cycles, it is possible that the initial entry site $e_0$ coincides with a pivot $\pi_n$ at some stage in the motion of the head of the worm. This can only happen if one reaches $\pi_n \equiv e_0$ via the only loop segment that touches $e_0$. {\em In this case, we randomly choose with probability $1/z$ each (where $z$ is the coordination number of the lattice) one of the neighbors of $\pi_n$ as the exit $e_n'$ (with the same operational meaning as before for ``entering a pivot'' and ``exiting a pivot'') and continue the worm construction}. In other words, the sole loop segment touching $e_0$ is moved to link $\langle e_0 e_n' \rangle$ from its original location on link $\langle e_0 e_n \rangle$, and the worm construction continues. 

The key point here is that this totally unbiased treatment of all exit options when $\pi_n$ coincides with $e_0$ does not spoil the detailed balance that is achieved by choosing {\em at all other pivots} an exit leg using a table of conditional probabilities that account for changes to the Boltzmann weight due to entering the pivot from a certain entry and exiting by a certain exit. Indeed, our procedure for randomly choosing the exit $e_n'$ whenever the current pivot coincides with $e_0$ merely adds an extra ``myopic step''~\cite{Kundu_Damle_PRX_2025,Kundu_Damle3Dpreprint2025,Rakala_Damle_2017,Morita_Lee_Damle_Kawashima2023}  over and above the ``myopic'' procedure already employed in the choice of the next pivot $\pi_{n+1}$ after an exit $e_n'$ is identified using the conditional probability table mentioned above. 

Here, by ``myopic'',  we mean a completely random step in the motion of the worm, which is therefore blind at that step to the resulting changes in Boltzmann weight. And indeed, as has been emphasized earlier~\cite{Rakala_Damle_2017,Morita_Lee_Damle_Kawashima2023} in other contexts as well, having such myopic steps interspered with steps that are sensitive to changes in the Boltzmann weight is actually one of the ways in which one can quite generally obtain a rejection-free worm algorithm that obeys detailed balance in many situations. It is thus clear that the original proof of detailed balance given in Ref.~\cite{Kundu_Damle3Dpreprint2025} clearly remains valid with this modification introduced here to deal with the nonbipartite case.

\subsection{Unit-vortex update}
Following the procedure introduced in Ref.~\cite{Kundu_Damle_PRX_2025}, we ignore for the purposes of this update all sites touched by nontrivial loop segments, as well as links connecting them to the remainder of the lattice. This yields a depleted lattice that is fully covered by dimers. On this depleted lattice, the standard dimer model worm algorithm~\cite{Alet_etal_PRE_2006,Sandvik_Moessner_2006} is used to generate a new fully-packed dimer configuration. Thus, this update leaves undisturbed all the nontrivial loops and only updates the dimers. As a result of this restricted mandate for this algorithm, the histogram of the head-to-tail distance of the worms of the dimer worm algorithm does not in this case directly yield the corresponding correlation function between the test $\delta S^z = \pm 2$ defects at the head and tail of the worm. This is a key difference from more standard settings in which the dimer worm algorithm has been used previously to obtain such information about correlations between test objects~\cite{Alet_etal_PRE_2006,Sandvik_Moessner_2006}.

\subsection{Technical details of simulations}
In practice, we used the ``heatbath'' versions of the probability tables that determine the choice of exit at each pivot. One Monte Carlo sweep or step consists of a certain number of worms of each type, such that each type of worm update results on average in an attempted update of ${\mathcal O}(L^2)$ link variables per step. To fix this number, we follow the following procedure: We first construct one worm of each type $20,000$ times and then, for the next $50,000$ such steps (comprising one worm of each type), we record the ``length'' of each worm (defined to be the number of pivot steps in its construction). From this we estimate the mean length of each worm for that value of $w$. Using this, we fix the number of worms of each type that constitute one Monte Carlo sweep or step.

With the definition of one Monte Carlo step fixed in this way, we perform a number of independent Monte Carlo runs (with different random number seeds) for a given $w$. Each run consists of a warm-up phase during which no measurements are made, and then a measurement phase in which we measure all observables after each Monte Carlo sweep, typically resulting in between $100,000$ to $250,000$ measurements in a given run at the larger sizes.  At these larger sizes, $5$ to $10$ such independent runs were used at each $w$. The final results quoted are based on an appropriately weighted average over all these independent Monte Carlo runs.  For the smaller sizes, a single long run comprising $10^6$ measurements sufficed.

\subsection{Measurements}\label{subsec:measurements}
After the initial warm-up phase of the Monte Carlo run, measurements of certain observables were performed after each Monte Carlo sweep. These are summarized below.

The most basic characteristic of a configuration is the density of sites touched by dimers, $\rho_d$, and the density of sites touched by non-trivial loops, $\rho_l$, normalized such that $\rho_d + \rho_l = 1$, independent of $w$. These are defined in the obvious way:
\begin{eqnarray}
\rho_{l} &=& \frac{1}{N_{\rm sites}} \sum_{j \in {\rm loops}} \langle s_j \rangle\nonumber \\
\rho_{d} &=& \frac{1}{N_{\rm sites}} \sum_{j \in {\rm dimers}} \langle s_j\rangle \; ,
\end{eqnarray}
where the angular brackets denote the Monte Carlo average, the first sum over $j$ is over all nontrivial loops and the second sum over $j$ is over all dimers in a configuration.
In addition, we keep track of the mean square fluctuation of $N_d$, the total number of dimers in a configuration. Apart from a prefactor, the magnitude of these mean-square fluctuations measures the specific heat of the corresponding frustrated magnet in the large $J/T$ limit (with $\mu/T$ held fixed) in which it maps to the dimer-loop model:
\begin{equation}
    \frac{c_v}{\log^2 w} = \frac{1}{L^2}\,\langle (\delta N_{d})^2 \rangle ,
    \label{eq:specificheat}
\end{equation}

Apart from measuring these densities and the specific heat $c_v$, we also measure the loop-size distribution $P_l(s)$ and the moments of the loop-size distribution, defined as
\begin{equation}
S_m \equiv \sum_{j=1}^{N_l} s_j^m, \quad m = 2,4,
\end{equation}
with the size of a loop $j$, denoted by $s_j$, defined as in the discussion in Sec.~\ref{subsec:Model} to be the total number of sites it passes through.  

From these moments, we constructed the loop-size Binder ratio
\begin{eqnarray}
 R &=& \langle S_4 \rangle /\langle S_2\rangle^2 \; ,
\end{eqnarray}
where the angular brackets denote Monte Carlo averaging over samples.
 In addition, we also keep track of the off-diagonal analogue of the Binder ratio, defined as:
\begin{eqnarray}
\mathcal{Q}_2 &= & \sum_{i \neq j}\langle s_i^2 s_j^2\rangle / \langle S_2\rangle^2  \; ,
\end{eqnarray}
where the angular brackets have the same meaning. 

These two quantities provide geometric information about the nature of the equilibrium state. In addition, we also measure the probability $P(W_x,W_y)$ that a configuration has $Z_2$-valued winding numbers $W_x$ and $W_y$ (defined in the discussion below Eq.~\eqref{eq:WindingNumberDef}), and use it to obtain the probability $P_{\mathrm{frac}}$ that at least one of these $Z_2$-valued winding numbers is $1$:
\begin{equation}
P_{\rm frac} = 1 - P(0,0)\; .
\end{equation}
This provides global topological information about the nature of the equilibrium state.

In addition, we recorded the histograms of the head to tail separation $\vec{r}$ of worms created by the half-vortex worm update and the unit-vortex worm update starting from an equilibrium ensemble of configurations. We denote these histograms by
 $h^{(q)}(\vec{r})$ with $q = 1/2, 1$ respecitvely to facilitate easy comparison with the results of Ref.~\cite{Kundu_Damle_PRX_2025} and Ref.~\cite{Kundu_Damle3Dpreprint2025}. We define these histograms as a count of the number of times (measured {\em per worm}) the displacement vector from the location of the tail of the worm to the location of the head of the worm is equal to $\vec{r}$. Here, the location of the tail is of course fixed at the original entry site $e_0$, while the location of the head of the worm is defined as the {\em location of the exit site $e_n'$ recorded after passing through each pivot $\pi_n$} (see Sec.~\ref{sec:Methods}).

Although our notation for these histograms is chosen to facilitate comparision with the corresponding results in the bipartite case~\cite{Kundu_Damle_PRX_2025,Kundu_Damle3Dpreprint2025}, we emphasize again that {\em there is no notion of vorticity for dimer-loop models on nonbipartite lattices}. Indeed, as we detail below, $h^{(q)}(\vec{r})$ is actually related to the correlation function of test excitations that have $\delta S^z_{\rm tot} = \pm 2q$ in the corresponding frustrated magnet.
 
From the point of view of the frustrated magnet, our results characterize the limiting behavior in the limit $J/T \rightarrow \infty$ with $\mu/T$ held fixed at a finite value. In this limit, each simplex of the frustrated magnet has $S^z_{\rm tot} = \alpha$ ($\alpha =2$ for the frustrated magnet that maps to the kagome lattice dimer-loop model, and $\alpha=4$ for the frustrated magnet that maps to the triangular lattice dimer-loop model). The histogram $h^{(1/2)}(\vec{r})$ serves as the Monte Carlo estimator for the correlation function $C^{(\pm 1)}(\vec{r})$ of two test defects, one of which has $\delta S^z_{\rm tot} = +1$ on one simplex and the other has $\delta S^z_{\rm tot} = -1$ on another simplex. 

This correlation function is defined as follows:
\begin{equation}
C^{\pm 1}(\vec{r}) = \frac{Z_{\pm 1}(\vec{r}, w)}{Z(w)},
\end{equation}
where $Z_{\pm 1}(\vec{r}, w)$ is the partition function of the fully packed dimer--loop model with two defects (corresponding to the two simplices with $\delta S^z_{\rm tot} = \pm 1$) separated by displacement $\vec{r}$, and $Z(w)$ is the corresponding partition function in the absence of defects. Note that by this definition $C^{\pm 1}(\vec{r} = 0) = 1$ since the numerator is identical to the denominator when $\vec{r} = 0$.

The argument that establishes that $h^{(1/2)}(\vec{r})$ is an unbiased Monte Carlo estimator of this correlation function is identical to the original arguments for the worm algorithm estimator of the correlation function of test defects given in earlier work~\cite{Alet_etal_PRE_2006,Sandvik_Moessner_2006,Alet_Sorensen_2003,Alet_Sorensen2003B}, and we do not repeat it here. Note that our definition and algorithmic procedure does guarantee that $h^{(1/2)}(\vec{r}=0) = 1$ as it should for this correspondence to hold.

However, the interpretation of the histogram $h^{(1)}(\vec{r})$ of head to tail separations $\vec{r}$ of worms constructed by the unit-vortex worm update is not as direct. Unlike in the previous case, this histogram cannot directly be used as a Monte Carlo estimator of the corresponding correlation function $C^{(\pm 2)}(\vec{r})$ of two $\delta S^z_{\rm tot} = \pm 2$ test defects in the frustrated magnet. This is due to the algorithmic restrictions involved in our construction of worms in the unit-vortex worm update, specifically, the fact that we do not allow the $\delta S^z_{\rm tot} = -2$ defect at the head of the worm split into two $\delta S^z_{\rm tot} = -1$ defects and later recombine before annihilating the $\delta S^z_{\rm tot} = +2$ defect that is held fixed at the fixed tail of the worm. 

Nevertheless, as noted in Ref.~\cite{Kundu_Damle_PRX_2025} and Ref.~\cite{Kundu_Damle3Dpreprint2025} for bipartite dimer-loop models, and in closely related work on topologically-distinct Coulomb phases of higher spin pyrochlore magnets for the analogous quantities~\cite{Pandey_Damle_Spin1preprint2025,Pandey_Kundu_Damle_Spin3by2preprint2026}, one can still obtain {\em partial information} about the long-distance behavior of $C^{(\pm 2)}(\vec{r})$ from the long distance behavior of $h^{(1)}(\vec{r})$: If $h^{(1)}(\vec{r})$ shows deconfinement, then it certainly implies that the actual defect correlator $C^{(\pm 2)}(\vec{r})$ is also deconfined. However, nothing definite can be concluded about the long-distance behavior of $C^{(\pm 2)}(\vec{r})$ if $h^{(1)}(\vec{r})$ is short-ranged, {\em i.e.}, shows confinement.

Finally, we flag a small detail about our conventions for defining the displacement vector $\vec{r}$ in the kagome case: In our actual measurements of these histograms, we define $\vec{r}$ to be the displacement vector from a reference site of the unit cell containing the tail of the worm to the corresponding reference site of the unit cell containing the head of the worm. For the triangular lattice, this is just the usual displacement vector between the two vertices of the triangular lattice (since the unit cell has just one site), and its definition then immediately implies that $h^{(1/2)}(\vec{r} = 0)=1$. However, for the kagome lattice, $h^{(1/2)}(\vec{r})$, as defined by us, is actually the estimator for the corresponding correlation function $C^{(\pm 1)}(\vec{r})$ summed over all $\vec{r}$ corresponding to a single unit cell of the kagome lattice; as a result of this definition $h^{(1/2)}(\vec{r} = 0) > 1$ in the kagome case. Similarly for $h^{(1)}(\vec{r})$.

\section{Results}
\label{sec:Results}

\begin{figure}
\centering
\includegraphics[width=\linewidth]{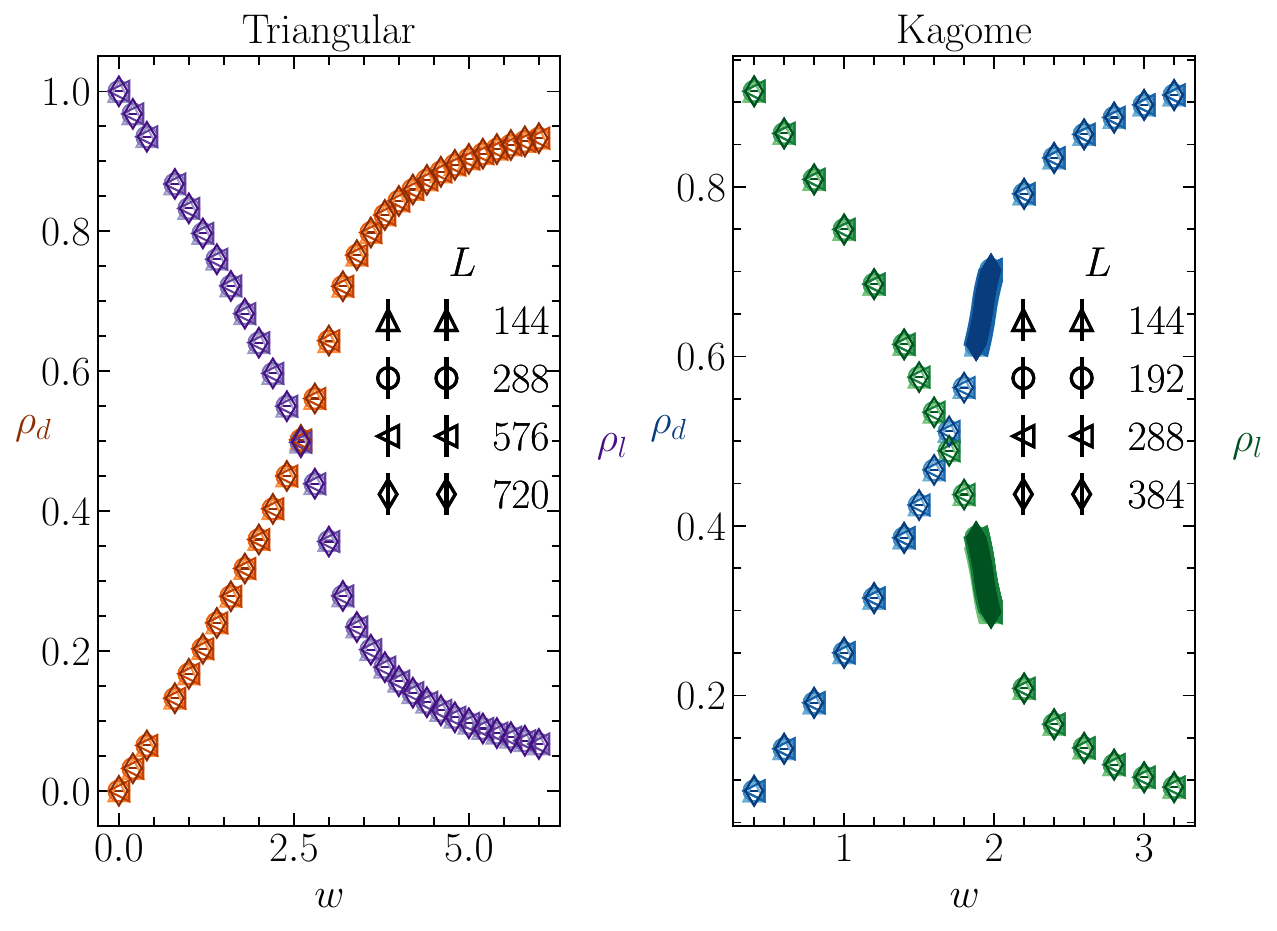}

\caption{The fraction of sites touched by dimers ($\rho_d$) and by loops ($\rho_l$) as a function of $w$ for both triangular and kagome lattice. These densities are nonzero for any finite $w >0$ and vary continuously with $w$.}
\label{fig:LoopDimerDensity}
\end{figure}

We begin by monitoring the mean density, $\rho_d$,  of sites touched by dimers (recall that the full packing constraint implies that the counterpart for loops, $\rho_l$, is just $1-\rho_d$ in our normalization). As we see in Fig.~\ref{fig:LoopDimerDensity}, both quantities remain nonzero for all nonzero finite values of $w$ and appear to vary smoothly with $w$. Nevertheless we now demonstrate that the system hosts two topologically-distinct liquid phases, separated from each other by a continuous phase transition at a finite critical value $w = w_c$. In addition, we present evidence that the transition is in the same Ising universality class as the flux confinement-deconfinement transitions identified recently in bipartite dimer-loop models~\cite{Kundu_Damle_PRX_2025,Kundu_Damle3Dpreprint2025}.

Indeed, the Binder ratios that are sensitive to the large-scale geometry of dimer-loop configurations show clear evidence of a sharply defined transition.  Consider for instance the Binder ratio
$\mathcal{Q}_2$. For small values of $w$ below a threshold, ${\mathcal Q}_2$ vanishes in the large-$L$ limit. This indicates that the geometry of dimer-loop configurations at small $w$ features large loops. On the other hand, for large $w$,  ${\mathcal Q}_2$  tends to a nonzero constant in the large-$L$ limit. This is a consequence of the fact that the configurations are made up of a mixture of extremely short loops and dimers (and thus continuously connected as far as their large scale structure is concerned with the infinite-$w$ limit, which is the fully-packed dimer model).
\begin{figure}
\centering
\includegraphics[width=\linewidth]{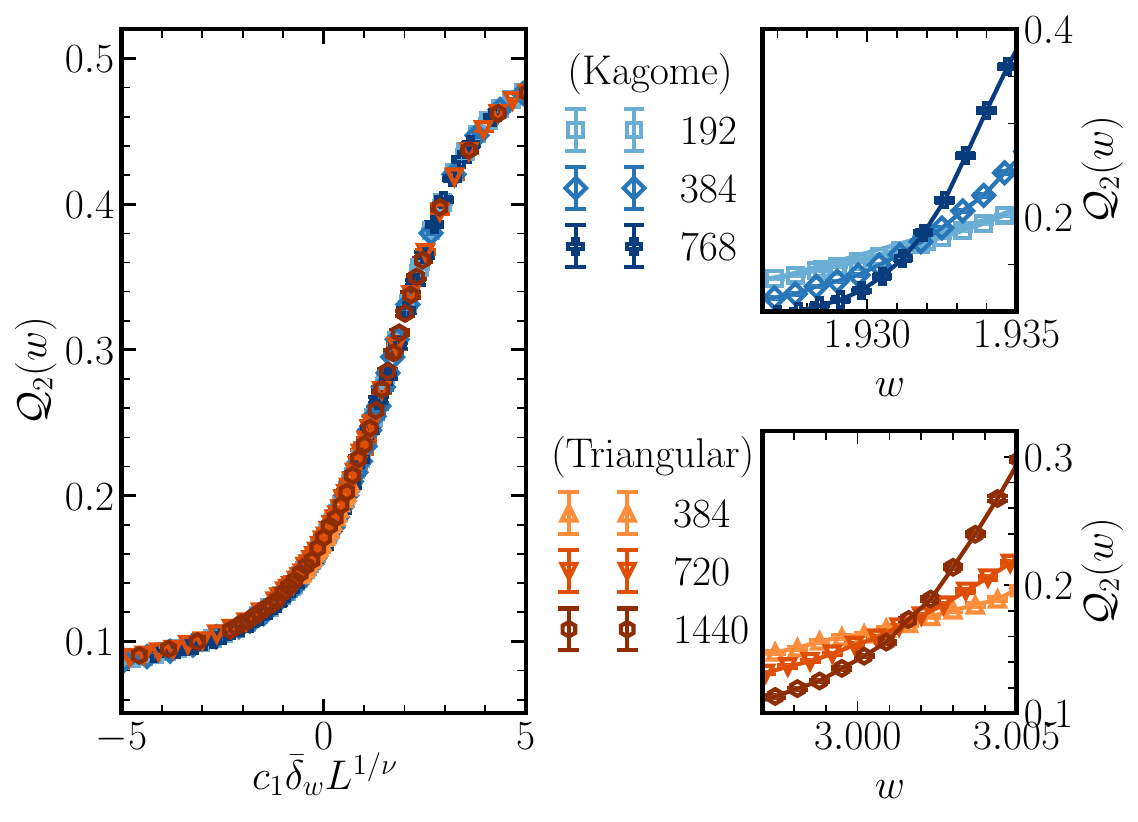}
\caption{Right: the Binder ratio $\mathcal{Q}_2$ of loop sizes defined in Sec. \ref{subsec:measurements} shows a clear crossing at critical value $w_c \approx 1.931(1)$ ($w_c\approx 3.001(1)$) on the kagome (triangular) lattice. Left: data for $w$ close to $w_c$ for various sizes collapse onto the scaling form described in Eq.~\eqref{eq:Q2Scaling}. The scaling collapse displayed here employs the following parameter values: $w_c = 3.0015$ ($w_c = 1.9315$) for the triangular (kagome) lattice case, and $\nu=1.0$ for both cases. Note that the choice $c_1 =1$ for the triangular lattice case is a convention that defines the scaling function $F_{{\mathcal Q}}$. The kagome data collapses on this scaling curve for $c_1 \approx 1.91(4)$. From the quality of the collapse, we obtain the estimate $\nu \approx 1.00(3)$.}
\label{fig:Q2}
\end{figure}

The distinction between these two regimes of behavior is sharp in the following sense: In the right panel of  Fig.~\ref{fig:Q2}, we see that data curves (corresponding to different system sizes) of the $w$ dependence of $\mathcal{Q}_2$ exhibit a clear and sharply defined crossing point at a critical value $w_c=3.001(1)$ for the triangular lattice and at $w_c=1.931(1)$ for the kagome lattice. Moreover, near the critical point, the data corresponding to different system sizes on both lattices collapse onto a universal critical scaling form
\begin{eqnarray}
    {\mathcal Q}_2(w,L) = F_{\mathcal Q}\!\left(c_1\,\bar{\delta}_w\, L^{1/\nu}\right) \label{eq:Q2Scaling}
\end{eqnarray} with $\bar{\delta}_{w}=(w-w_c)/w_c$ and $\nu=1.00(3)$. Here $c_1$ is a lattice-dependent scale factor defined to be $1$ on the triangular lattice and estimated to be $c_1=1.91$ for the kagome lattice. The high quality of this scaling collapse can be seen in the left panel of Fig.~\ref{fig:Q2}. For completeness, we also note that the other Binder ratio $R$ also shows an analogous signature of the transition at a value of $w_c$ that is consistent with the estimates from these results for ${\mathcal Q}_2$ (not shown here). Thus we conclude that these nonbipartite dimer loop models have a sharp transition from a small-$w$ long-loop phase to a large-$w$ short-loop phase. This behavior is very similar to the corresponding transition in the bipartite dimer-loop models studied in Ref.~\cite{Kundu_Damle_PRX_2025} on the square and honeycomb lattices.

Next, we turn to a study of the $Z_2$-valued winding numbers $W_x$ and $W_y$ that characterise topological aspects of the dimer-loop configurations, specifically focusing on $P_{\rm frac}$, the probability that at least one of them is $1$, equivalently, the probability that at least one of $\phi_x$ and $\phi_y$ has a fractional part of $1/2$. This is an indication of the presence of long system spanning loops, since a single loop that wraps around a periodic direction gives rise to a fractional part of $1/2$ to the transverse component of $\vec{\phi}$, while short loops only lead to integer contributions to $\phi_x$ and $\phi_y$.
\begin{figure} 
\centering
\includegraphics[width=\linewidth]{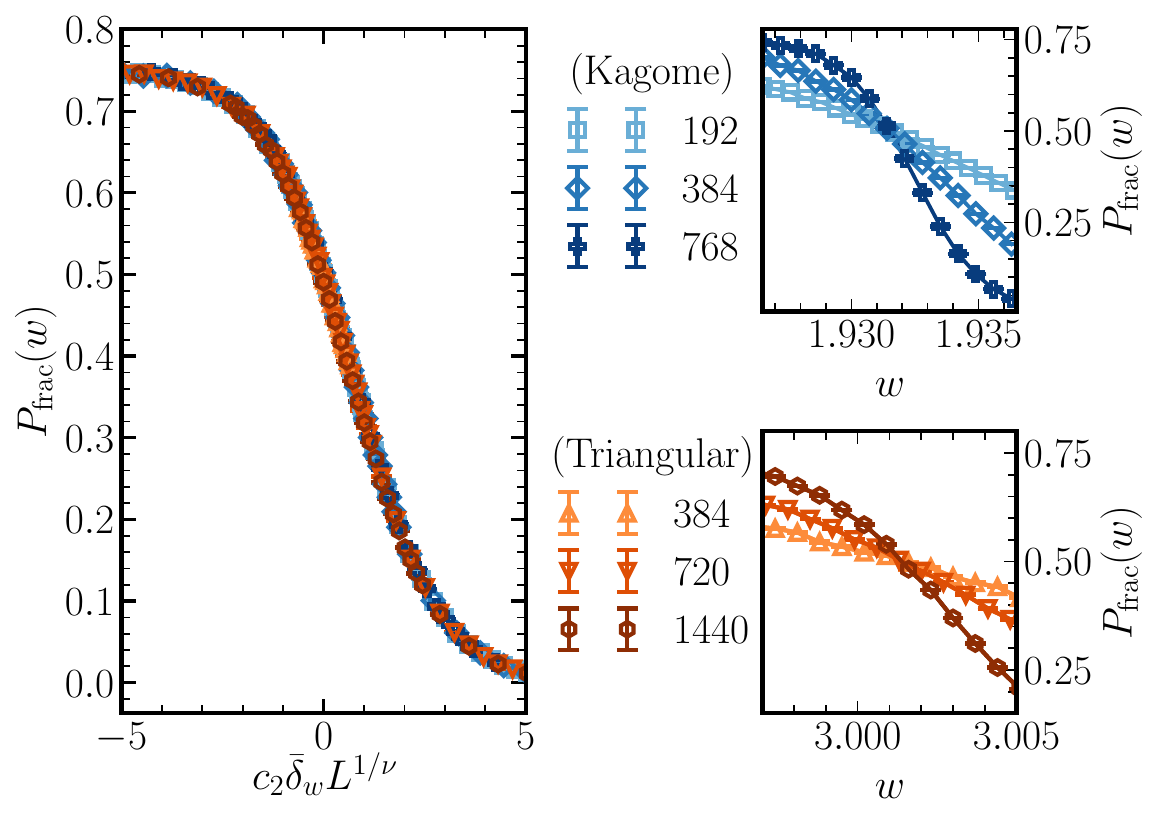}
\caption{Right: the probability of finding fractional winding (either $W_x=1$ or $W_y=1$ or both)  defined in Sec. \ref{subsec:measurements}, shows a clear crossing at the critical value $w_c \approx 1.931(1)$ ($w_c\approx 3.001(1)$) on the kagome (triangular) lattice. Left: data for $w$ close to $w_c$ for various sizes $L$ collapse onto the scaling form described in Eq.~\eqref{eq:PScaling}. The scaling collapse displayed here employs the following parameter values: $w_c = 3.0015$ ($w_c = 1.9315$) for the triangular (kagome) lattice case, $\nu=1.0$ for both cases. Note that the choice $c_2 =1$ for the triangular lattice case is a convention that defines the scaling function $F_{P}$ from the collapsed triangular lattice data. We find that the kagome data collapses on to this scaling curve for $c_2 \approx 1.95(3)$. From the quality of the scaling collapse, we obtain the estimate $\nu \approx 1.00(3)$ on both lattices.}
\label{fig:Pfrac}
\end{figure}

We find that $P_{\rm frac}$ tends to a nonzero constant in the large-$L$ limit at small $w$. This nonzero probability is close to $3/4$ in value. One can understand this roughly to mean that all four $Z_2$ winding number sectors are equally probable in the small-$w$ phase (since only one of the four sectors is characterized by both $W_x$ and $W_y$ being equal to $0$). On the other hand, at large $w$, we find that $P_{\rm frac}$ vanishes in the large $L$ limit. 

The fact that the entire large-$w$ phase is characterized by a vanishing $P_{\rm frac}$ in the large $L$ limit is surprising in the following sense: Loops are strictly forbidden only at $w=\infty$. One might therefore have expected that the configurations at any finite $w$, however large, would still have the occasional large loop, and this would be reflected in a small but nonzero value of $P_{\rm frac}$. The fact that this does not happen is an interesting entropic effect that stabilizes the large-$w$ phase, not an energetic prohibition (as it is at $w=\infty$).

The right panel of Fig.~(\ref{fig:Pfrac}) shows the $w$ dependence of $P_{\rm frac}$ for different system sizes on both lattices. We see that the data exhibits a sharp crossing of these curves, and the value of $w$ at the crossing point is within the errors of the previous estimate for $w_c$ (from the loop-size Binder cumulants) for each lattice. Moreover, from the left panel of this figure, we see that the data for different system sizes on both lattices collapse onto a single curve, indicating that $P_{\rm frac}(w,L)$ obeys the following scaling form at large $L$ near $w_c$:
\begin{equation}
P_{\rm frac}(w,L) = F_{P}\!\left(c_2\bar{\delta}_w L^{1/\nu}\right),
\label{eq:PScaling}
\end{equation}
where $\bar{\delta}_w = (w - w_c)/w_c$, $\nu = 1.00(1)$, and $c_2$ is a nonuniversal, lattice-dependent scale factor defined to be $1$ on the triangular lattice and estimated to be $\approx 1.95$ for the kagome lattice. 
\begin{figure}
\centering
\includegraphics[width=\linewidth]{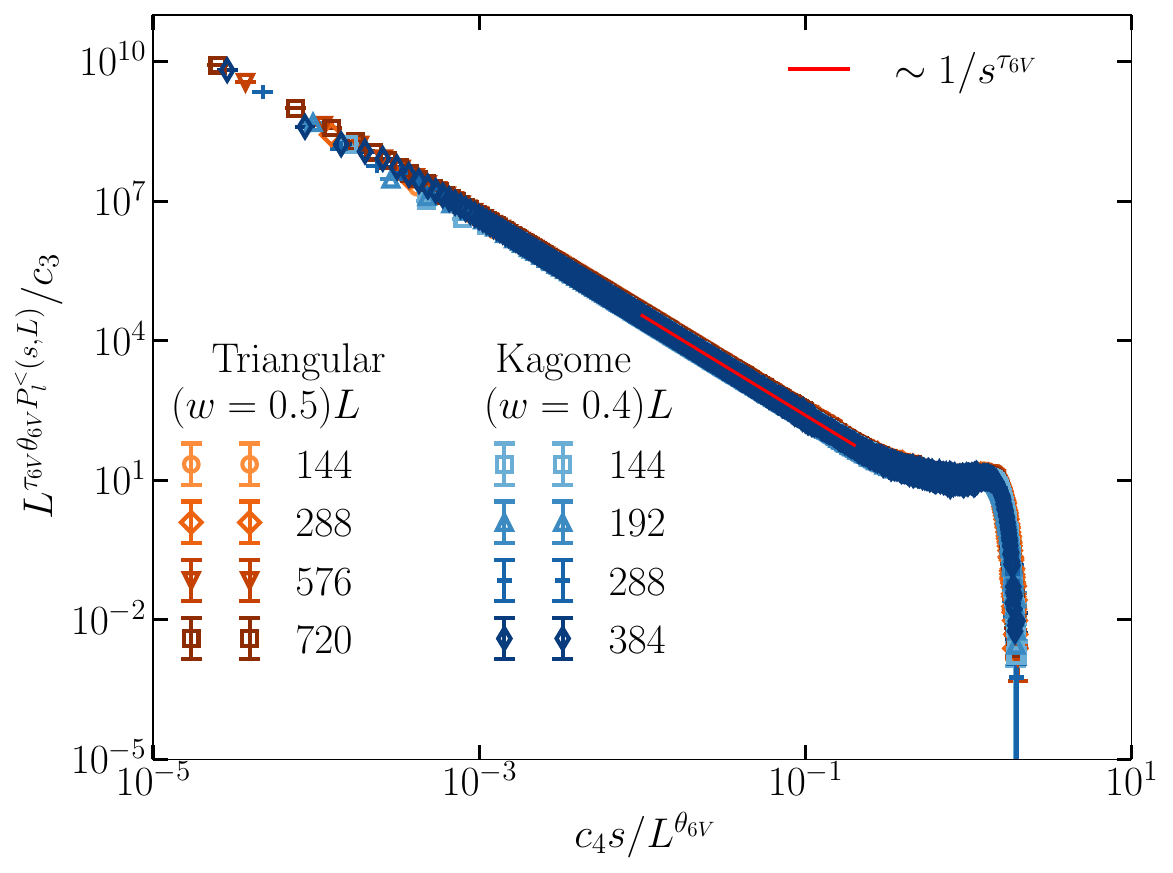}
\caption{Data for the loop size distribution $P^{<}_{l}(s,L)$ for various sizes $L$ for all
$w<w_c$ obeys the scaling form defined in Eq.~\eqref{eq:LongLoopDist}, with exponents $\theta_{6V}$ and $\tau_{6V}$ corresponding to the exactly known values for the six-vertex model on the square lattice. This has been illustrated above with two representative $w$ values: $w = 0.5$ on the triangular lattice and $w=0.4$ on the kagome lattice. The choice of $c_3=c_4=1$ for the triangular lattice at $w=0.5$ is a convention. With this convention, we find that the kagome lattice data for $w=0.4$ collapses onto the same curve when we choose $c_4 \approx 0.38(1)$ and $c_3 \approx 0.13(1)$.}
\label{fig:LongLoopDistribution}
\end{figure}

Next, we study the loop-length distribution $P_l(s,L)$ in the $w < w_c$ long-loop phase. For all $w < w_c$ on both lattices, we find that $P_l(s,L)$ for different system sizes collapses onto a universal finite-size scaling form
\begin{equation}
P_l^{<}(s,L) = \frac{c_3}{L^{\tau_{6V}\theta_{6V}}}
G_{6V}\!\left(\frac{c_4 s}{L^{\theta_{6V}}}\right),
\label{eq:LongLoopDist}
\end{equation}
where $\tau_{6V} = 15/7$ and $\theta_{6V} = 1.75$ are characteristic exponents of the exactly solvable six-vertex model~\cite{Jaubert_Haque_Moessner_2011,Saleur_Duplantier_1987,Kondev_Henley_1995,Jacobsen_Kondev_1998}, and $c_3$ and $c_4$ are nonuniversal scale factors that depend on the lattice and on $w$. In Fig.~\ref{fig:LongLoopDistribution}, we demonstrate this scaling for several representative values of $w$ on both lattices.

In the long-loop phase, this scaling behavior implies that the size $s$ of the largest loop in a sample of size $L$ scales as $s \sim L^{\theta_{6V}}$~\cite{Kundu_Damle_PRX_2025}. For loop sizes in the range $1 \ll s \ll L^{\theta_{6V}}$, this scaling also implies~\cite{Kundu_Damle_PRX_2025} that the distribution has power-law behavior $P_l^{<}(s) \sim s^{-\tau_{6V}}$. Finally, as noted in Ref.~\cite{Kundu_Damle_PRX_2025}, the fact that the exponents obey $\tau_{6V}\theta_{6V} = 2 + \theta_{6V}$ implies that there are only $\mathcal{O}(1)$ loops of this maximal size $\sim L^{\theta_{6V}}$ in a typical dimer-loop configuration in the long-loop phase. All of this is entirely analogous to the behaviour found in Ref.~\cite{Kundu_Damle_PRX_2025} in the bipartite case, and matches the expected behavior of loops in the exactly solvable six-vertex model on the square lattice.

\begin{figure} 
\centering
\includegraphics[width=\linewidth]{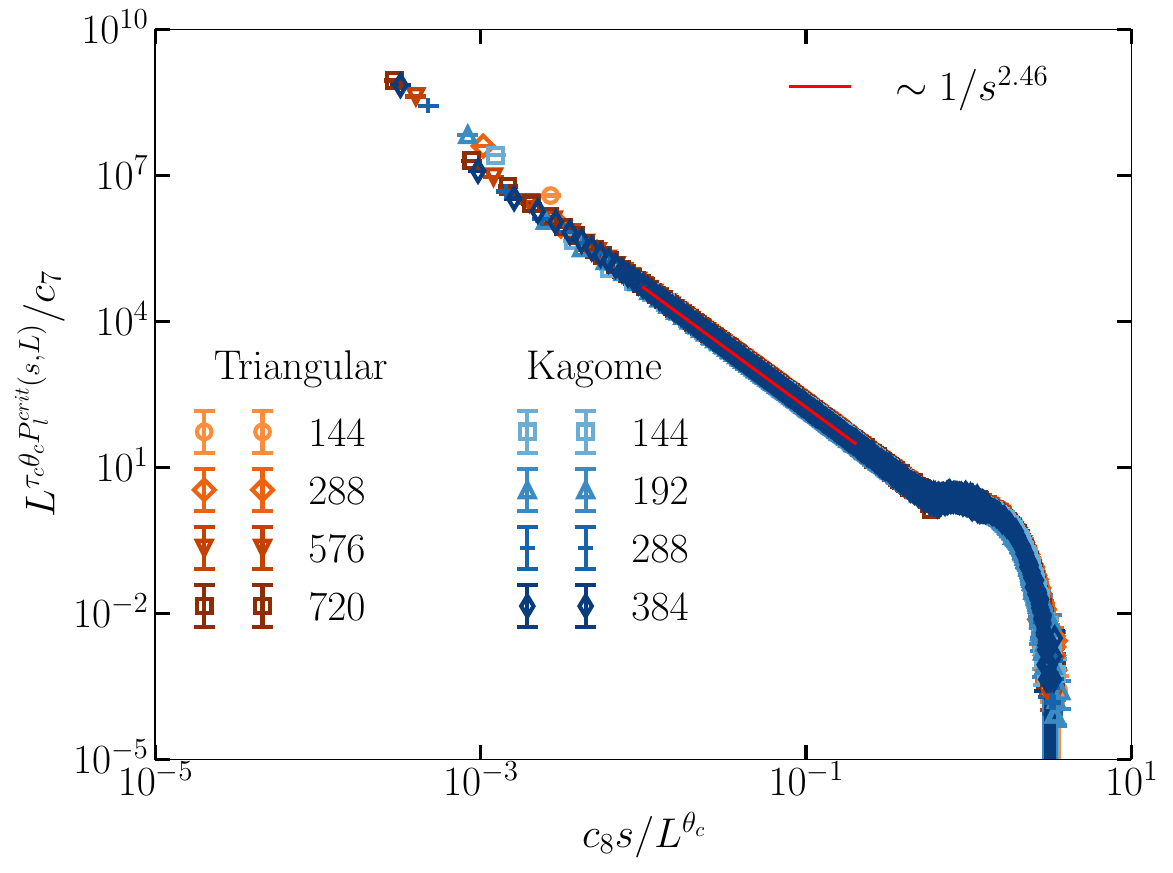}
\caption{Data for the critical loop size distribution $P^{\rm crit}_{l}(s,L)$ at $w_c=3.001$ ($w_c=1.931$) for the triangular (kagome) lattice for various sizes $L$ on both lattices obeys the scaling form defined in Eq.~\eqref{eq:Pcrit} with common exponent choices $\theta_c = 1.37$ and $\tau_c = 2.46$ for both lattices. As a matter of convention, we take $c_8 = c_7 =1.0$ for the triangular lattice case.  With this convention we find that the kagome data collapses onto the same scaling curve for  $c_8 \approx (0.45,0.47)$ and $c_7 \approx (0.14,0.16)$. From the quality of the collapse, we estimate $\theta_c \approx 1.37(6)$ and $\tau_c \approx 2.46(10)$.}
\label{fig:CriticalLoopDistribution}
\end{figure}

The critical loop-length distribution on both lattices, shown in Fig.~\ref{fig:CriticalLoopDistribution}, collapses onto a different universal finite-size scaling form,
\begin{equation}
P_l^{\rm crit}(s,L) = \frac{c_7}{L^{\tau_c \theta_c}}
G_{\rm crit}\!\left(\frac{c_8 s}{L^{\theta_c}}\right),
\label{eq:Pcrit}
\end{equation}
similar to that in the long-loop phase, but with a different set of universal exponents, $\theta_c = 1.37(6)$ and $\tau_c = 2.46(10)$, which again satisfy $\tau_c \theta_c = 2 + \theta_c$ within statistical error. Here, $c_7$ and $c_8$ are lattice-dependent non-universal scale factors defined to be $1$ on the triangular lattice and taking on values $c_8 \approx 0.46$ and $c_7 \approx 0.15$ for the kagome lattice case. This scaling now implies that i) the size of the largest loops at the critical point scales as $s \sim L^{\theta_c}$, ii) that there are only $\mathcal{O}(1)$ such large loops in a configuration at criticality, and iii) For $1 \ll s \ll L^{\theta_c}$, the loop size distribution has power-law behavior $P_l^{\rm crit}(s) \sim s^{-\tau_c}$. Note that these values of $\theta_c$ and $\tau_c$ are consistent within errors with the corresponding values found at the critical point in the bipartite case in Ref.~\cite{Kundu_Damle_PRX_2025}. 

\begin{figure}
\centering
\includegraphics[width=\linewidth]{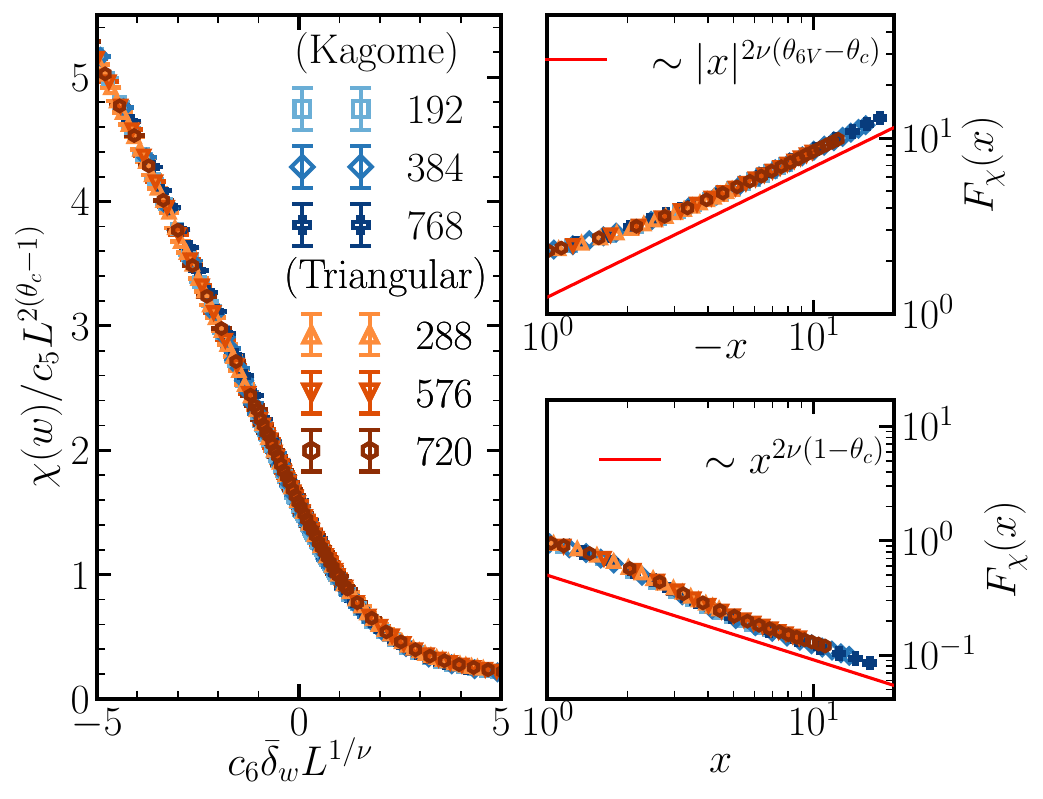}
\caption{Left: the loop size susceptibility on both the triangular and the kagome lattice is seen to obey the scaling form of Eq.~\eqref{eq:chiScalingForm}, with a common choice of $\theta_c=1.37$ and $\nu=1.0$ for the exponents, and $w_c=3.001$ ($w_c=1.931$) for the triangular (kagome) lattice. As a matter of convention, we set  $c_5 = c_6 =1$ for the triangular lattice case. With this convention, we find that the kagome lattice data collapse onto the same scaling curve for $c_5 \approx 4.70(5)$ and $c_6 \approx 1.90(5)$. Right: The top and bottom panels show the asymptotic behavior of $F_{\chi}(x)$ for $x\ll 0$ and $x\gg 0$ respectively. For $x\gg 0$, $F_{\chi}(x)\sim x^{-k}$, where $k=2\nu(\theta_c -1) \approx0.74(2)$, while for $x\ll 0$, $F_{\chi}(x)\sim |x|^{-p}$, where $p=2\nu(\theta_{6V} -\theta_c) \approx0.76(2)$. These data are consistent with the scaling of the loop size distribution at criticality and in both phases.}
\label{fig:Susceptibility}
\end{figure} 

This scaling behavior of the largest loops is also probed by the loop-size susceptibility $\chi$, defined as
\begin{equation}
\chi = \frac{\langle S_2\rangle}{L^2},
\label{eq:chi}
\end{equation}
where $S_2$ is the second moment of the loop-size distribution. From the loop-size distributions in the long-loop phase and at the critical point, we expect $\chi \propto L^{2(\theta_{6V}-1)}$ in the $w < w_c$ long-loop phase and $\chi \propto L^{2(\theta_c-1)}$ at the critical point. In contrast, in the short-loop phase, we expect $\chi$ saturates to an $\mathcal{O}(1)$ value, reflecting the predominance of dimers and short loops. Our data for $\chi$ is consistent with these expectations. Indeed, from Fig.~(\ref{fig:Susceptibility}), we see that our data for $\chi$ on both lattices collapses onto a universal scaling form
\begin{equation}
\chi = c_5 L^{2(\theta_c - 1)} F_{\chi}\!\left(c_6 \bar{\delta}_w L^{1/\nu}\right),
\label{eq:chiScalingForm}
\end{equation}
with $\nu = 1.00(3)$ and $\bar{\delta}_w = (w - w_c)/w_c$. Here, $c_5$ and $c_6$ are lattice-dependent scale factors defined to be $1$ on the triangular lattice and taking on values $c_5 \approx 4.7$ and $c_6 \approx 1.93$ on the kagome lattice.

Away from the critical point, the asymptotic behavior of the scaling function $F_{\chi}(x)$ is shown in the right panels of Fig.~\ref{fig:Susceptibility}. For $x \gg 0$, $F_{\chi}(x) \sim x^{-2\nu(\theta_c - 1)}$, whereas for $x \ll 0$, $F_{\chi}(x) \sim |x|^{2\nu(\theta_{6V} - \theta_c)}$. These limiting forms follow directly from the loop-size distributions in the two phases and at the critical point.

\begin{figure} 
\centering
\includegraphics[width=\linewidth]{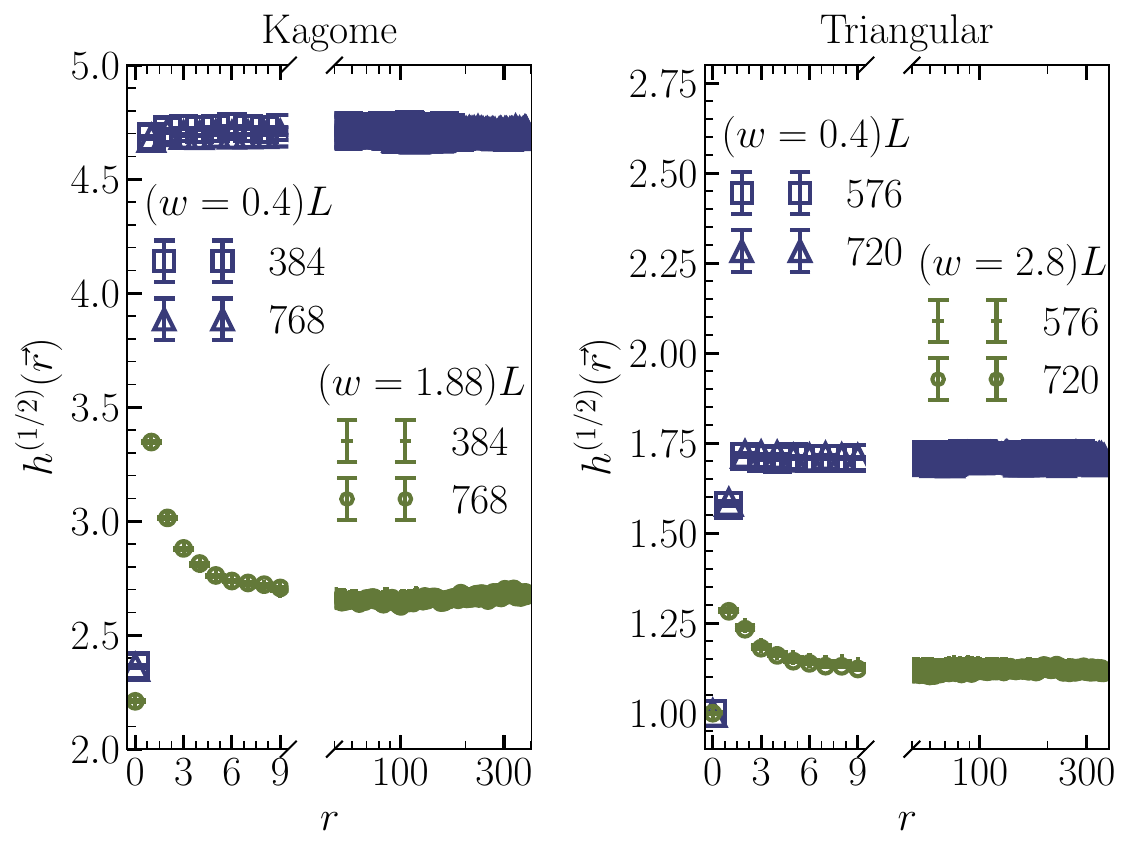}
\caption{The histogram $h^{(1/2)}(\vec r)$ of head-to-tail separations of half-vortex worms saturates to a constant non-zero value for $w<w_c$ on both the kagome lattice (left), and the triangular lattice (right).}
\label{fig:LongLoopHalfVortex}
\end{figure}

\begin{figure} 
\centering
\includegraphics[width=\linewidth]{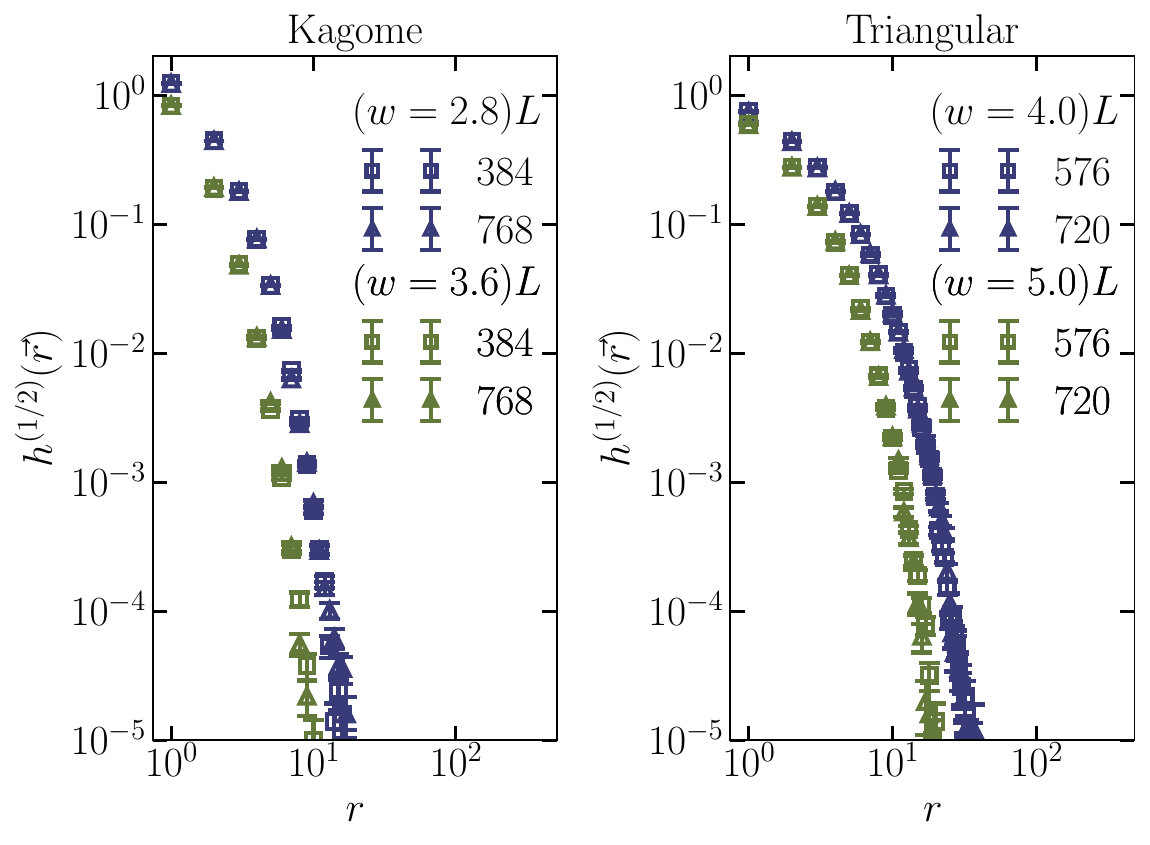}
\caption{The histogram $h^{(1/2)}(\vec r)$ of head-to-tail separations of half-vortex worms in the $w>w_c$ short-loop phase decays rapidly to zero, faster than any power-law, on both the kagome lattice (left), and the triangular lattice (right).}
\label{fig:ShortLoopHalfVortex}
\end{figure}

\begin{figure} 
\centering
\includegraphics[width=\linewidth]{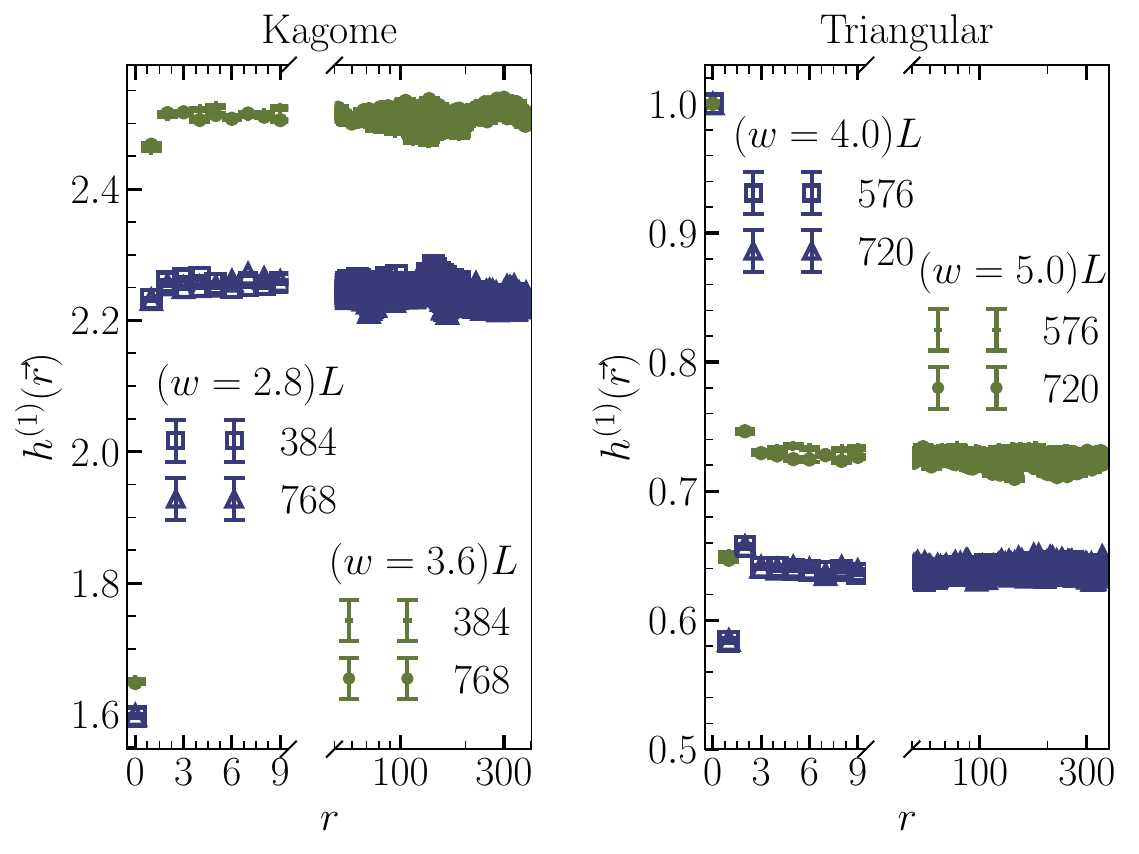}
\caption{The histogram $h^{(1)}(\vec r)$ of head-to-tail separations of unit-vortex worms saturates to a constant non-zero value for $w>w_c$ on both the kagome lattice (left), and the triangular lattice (right).}
\label{fig:ShortLoopUnitVortex}
\end{figure}

\begin{figure} 
\includegraphics[width=\linewidth]{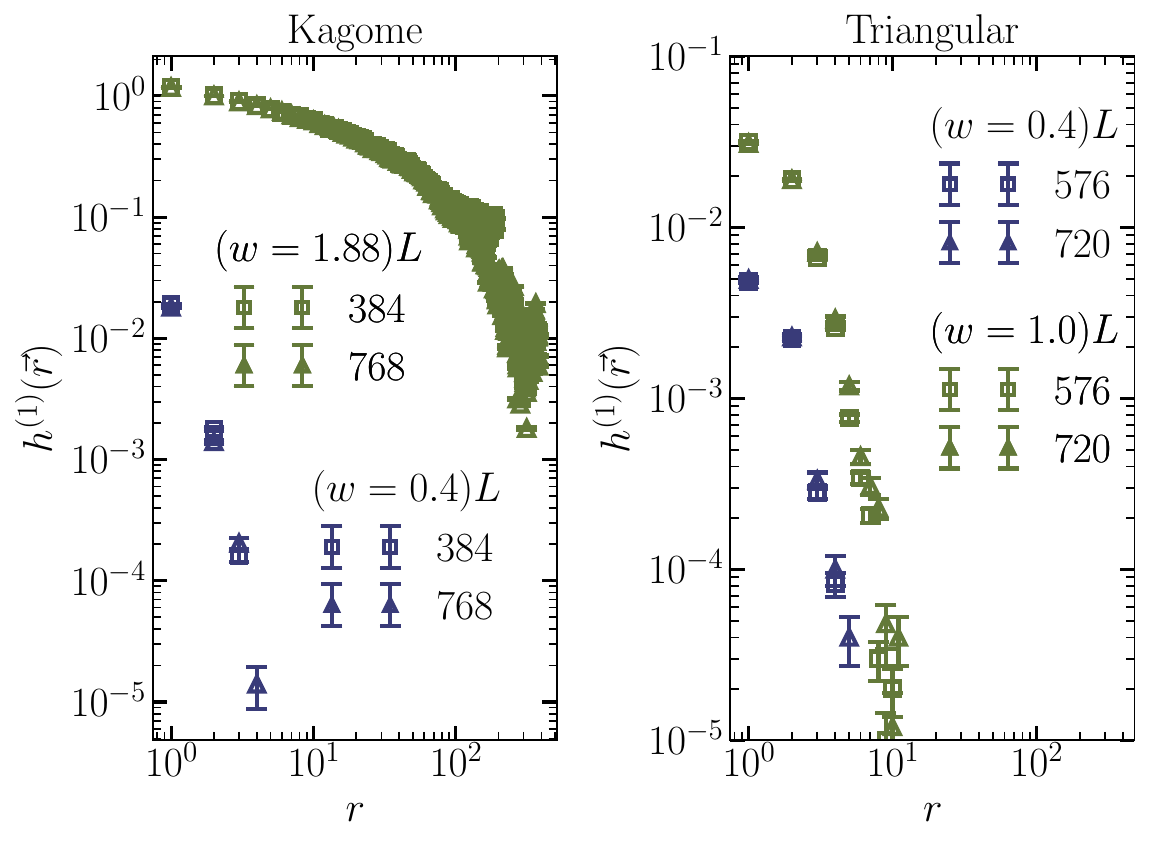}
\caption{The histogram $h^{(1)}(\vec r)$ of head-to-tail separations of unit-vortex worms in the $w<w_c$ long-loop  phase decays rapidly to zero, faster than any power-law, on both the kagome lattice (left), and the triangular lattice (right).}
\label{fig:LongLoopUnitVortex}
\end{figure}

\begin{figure}
\centering
\includegraphics[width=\linewidth]{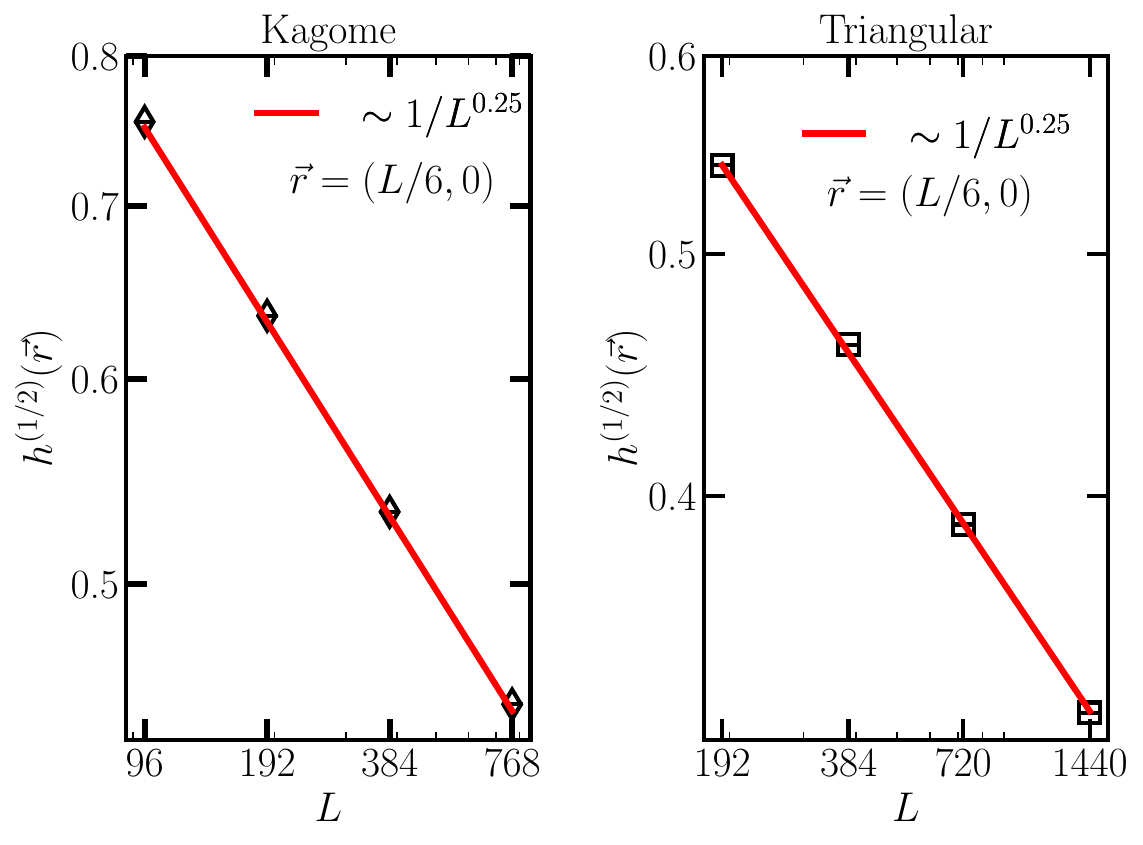}
\caption{On both the kagome and the triangular lattices, the histogram $h^{(1/2)}(\vec r)$ of head-to-tail separations of half-vortex worms in the vicinity of the critical point for $\vec{r} = (L/6,0)$ decays as a power-law. From this data, we estimate the exponent of this power-law to be $\eta^{(\pm 1)} \approx 0.250(4)$ on both lattices.}
\label{fig:CriticalHalfVortex}
\end{figure}

Next, we discuss the behavior of the histograms $h^{(1/2)}(\vec{r})$ and $h^{(1)}(\vec{r})$ of the head-to-tail displacement $\vec{r}$ of worms constructed by the half-vortex and 
unit-vortex worm updates starting from an equilibrium configuration. As noted earlier, $h^{(1/2)}(\vec{r})$ provides us a direct measurement of the test-defect correlator $C^{(\pm 1)}(\vec{r})$ for a
 pair of defects corresponding to $\delta S^z_{\rm tot} = \pm 1$ on two simplices of the corresponding frustrated magnet, while $h^{(1)}(\vec{r})$ gives partial information about the analogous correlator $C^{(\pm 2)}(\vec{r})$ for two $\delta S^z_{\rm tot} = \pm 2$ defects. 
  We find that $h^{(1/2)}(\vec{r})$ shows deconfined behavior (tends to a nonzero limit at large $r=|\vec{r}|$) throughout the small-$w$ long-loop phase, but is confined (decays rapidly to zero as $r$ is increased) for $w>w_c$. Thus, $C^{(\pm1)}(\vec{r})$ tends to a nonzero limit at large $r$ in the small-$w$ phase but is short-ranged in the large-$w$ phase. This is shown in Fig.\ref{fig:LongLoopHalfVortex} and Fig.~\ref{fig:ShortLoopHalfVortex}. On the other hand, we find that $h^{(1)}(\vec{r})$ is deconfined in the large-$w$ short-loop phase, implying that $C^{(\pm 2)}(\vec{r})$ is long-ranged, tending to a nonzero limit for large $r$ in this phase. This is shown in Fig.~\ref{fig:ShortLoopUnitVortex}. 
  
In the small-$w$ long-loop phase, we find that $h^{(1)}(\vec{r})$ is short-ranged, but, as mentioned earlier, this is an algorithmic artifact, and {\em does not} imply confinement for the correlator $C^{(\pm 2)}(\vec{r})$. Indeed, viewing a spin $|\delta S^z_{\rm tot}| = 2$ defect as a pair of $\delta S^z_{\rm tot}| = 1$ defects bound to each other, we expect that $C^{(\pm 2)}(\vec{r})$ is in fact also long-ranged in the small-$w$ long-loop phase, and does not see the transition between the two phases.

At the critical point, we find that $h^{(1/2)}(\vec{r})$ falls off as a power law at large $r$, with power-law exponent $\eta^{(\pm 1)}  \approx 0.250(4)$ within statistical errors, implying that $C^{(\pm 1)}(\vec{r}) \sim 1/r^{\eta^{(\pm 1)}}$. This value of $\eta^{(\pm 1)} = 1/4$ is precisely what one would predict at $w_c$ if the short-loop to long-loop transition is actually controlled by an Ising theory. To see this, we need to first understand the precise sense in which the transition is an Ising transition. 
This question is answered in parallel work~\cite{Pandey_Damle_DimerLoopIsingorderarXiv2026}. Here, we review the key idea~\cite{Pandey_Damle_DimerLoopIsingorderarXiv2026} that forms the basis of that work, and then use it to give a fairly precise scaling argument for the value of $\eta^{(\pm 1)}$ in the present case.

The key idea~\cite{Pandey_Damle_DimerLoopIsingorderarXiv2026} is this: The nontrivial loops of the dimer-loop model should be thought of as domain walls of an Ising spin field $\sigma$ defined on the planar dual of the lattice whose links host the dimers and the nontrivial loops. With this idea in hand, we can now interpret  histogram $h^{(1/2)}(\vec{r})$ (equivalently the test-defect correlator $C^{(\pm1)}(\vec{r})$) in terms of the physics of this Ising field $\sigma$. To do this we note that the worm creates a configuration of dimers and loops with an open string of loop segments connecting the head and tail. This maps precisely to a fractional domain wall in the Ising field $\sigma$, exactly as envisaged in the original Kadanoff-Ceva~\cite{Kadanoff_Ceva_1971,Fradkin_Review2017} definition of the Ising disorder operators $\tau$. Thus, $C^{(\pm1)}(\vec{r})$ is the correlator of the disorder field $\tau$ of this Ising theory. Since the Ising transition in two dimensions has self-dual character, critical correlations of $\tau$ are controlled by the anomalous exponent $\eta_{\rm Ising} = 1/4$, implying that
$C^{(\pm1)}(\vec{r}) \sim 1/r^{\eta_{\rm Ising}}$ at criticality. This provides a rather precise rationale for the observed value of $\eta^{(\pm 1)}  \approx 0.250(4)$.

\begin{figure} 
\centering
\includegraphics[width=\linewidth]{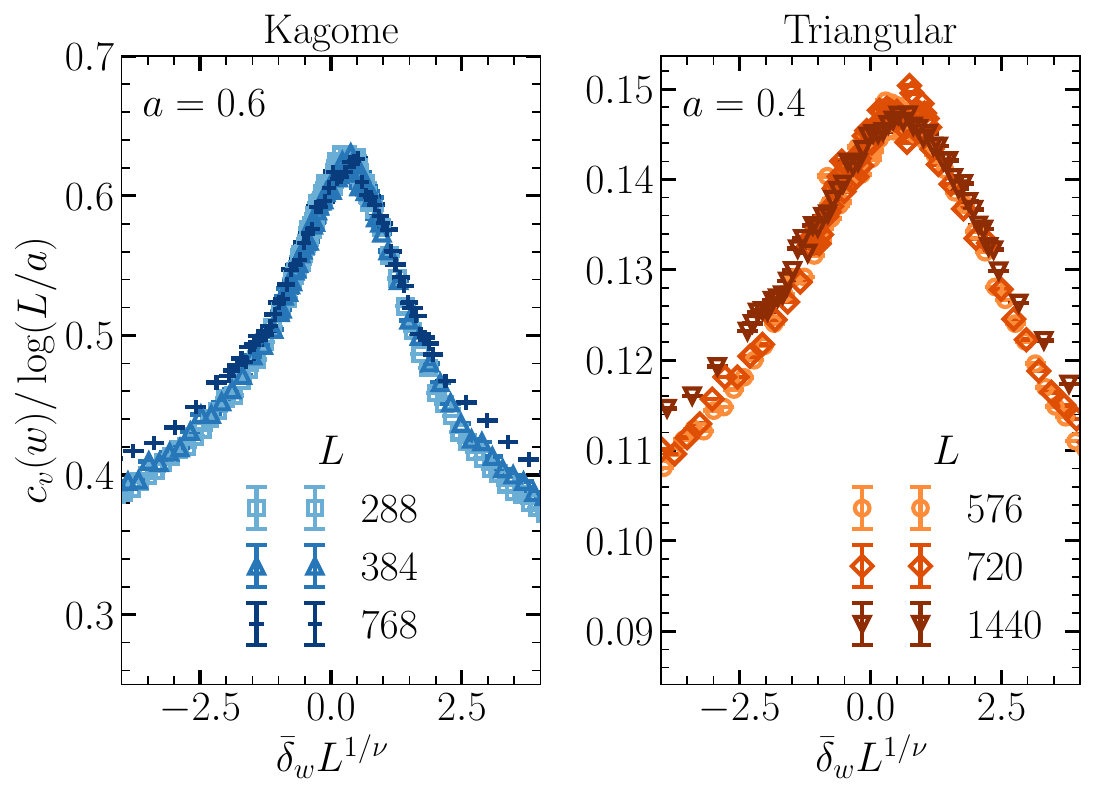}
\caption{The specific heat near the transition on both the kagome and triangular lattices exhibits a singular behavior that appears consistent with a logarithmic divergence, and data in the vicinity of the transition displays reasonably good scaling collapse with the correlation length exponent set to $\nu = 1.0$. 
%For the kagome lattice $a \approx (0.35,0.75)$, while for the triangular lattice $a \approx (0.2,0.5)$.}
}
\label{fig:SpecificHeat}
\end{figure}

The transition at $w_c$ also leaves a weak but discernible imprint on the behavior of the specific heat. We find that the $w$ and $L$ dependence of the specific heat in the critical region displays a finite-size scaling collapse. The quality of the scaling collapse can be seen in Fig.~(\ref{fig:SpecificHeat}). The singular behavior can be fit to a logarithmic divergence at $w_c$ on both lattices, consistent with Ising criticality, although a very weak power-law divergence cannot be conclusively excluded based on this observable alone.

\section{Discussion}
\label{sec:Discussion}
Thus, we have established that fully-packed dimer-loop models in two dimensions exhibit an interesting topological transition driven by the relative fugacity $w$ of dimers even when the lattice in question is not bipartite. Several of the universal critical properties of such transitions on non-bipartite two-dimensional lattices correspond quite precisely to the analogous properties at such transitions on bipartite lattices. Specifically, the correlation length exponent, and, equivalently, the nature of the singularity in the specific heat appear to be independent of the bipartiteness of the lattice and match the behavior expected in the Ising universality class in two dimensions. In addition, the loop size statistics in the long-loop phase and its critical behavior at the transition are both blind to the distinction between bipartite and nonbipartite lattices. 
This is to some extent surprising, given that the bipartite transition has an interpretation in terms of a confinement-deconfinement transition of a U($1$) flux degree of freedom, while the nonbipartite transition has to do with some $Z_2$ winding sectors being suppressed in the thermodynamic limit in one phase but not the other. 

However, other properties, in particular the power-law behavior of the histogram of head-to-tail separations of the half-vortex worm at criticality, and the behavior of this histogram in the long-loop phase are both quite different in the two cases. In addition, in the bipartite case, both phases feature power-law correlations of some local operators of the corresponding frustrated magnet, while this is not the case in the nonbipartite situation. 

Some aspects of these results are further clarified in parallel work~\cite{Pandey_Damle_DimerLoopIsingorderarXiv2026} that focuses on identifying a local Ising degree of freedom in such dimer-loop models and then understanding the properties of the phases and the transitions from this perspective. Other questions remain unanswered, for instance: Do non-bipartite dimer-loop models in three-dimensions also feature an Ising transition like their bipartite counterparts studied in Ref.~\cite{Kundu_Damle3Dpreprint2025}? We hope to return to such questions in future work.

\section*{Acknowledgements}
We gratefully acknowledge generous allocation of computing resources by the Department of Theoretical Physics (DTP) of the Tata Institute of Fundamental Research (TIFR), and related technical assistance from K. Ghadiali and A. Salve. The work of JP was supported at the TIFR by a graduate fellowship from DAE, India, while the work of SK at the International Center for Theoretical Sciences of TIFR (ICTS-TIFR)  was supported by a postdoctoral fellowship from DAE, India, under project no. RTI4019.  KD was supported at the TIFR by DAE, India, and in part by a J.C. Bose Fellowship (JCB/2020/000047) of SERB, DST India, and by
the Infosys-Chandrasekharan Random Geometry Center (TIFR).
% \nocite{*}
\bibliography{bibliographyMerged}% Produces the bibliography via BibTeX.

\end{document}